\documentclass[pre,
 reprint,
 superscriptaddress,
 nofootinbib,
 noeprint,
 amsmath,amssymb,
 aps,
 floatfix
]{revtex4-2}

\usepackage[textheight=672pt,textwidth=510pt]{geometry}

\usepackage[utf8]{inputenc}
\usepackage{graphicx}
\usepackage{amsmath,amssymb,amsfonts}
\usepackage{bm}
\usepackage{siunitx}

\usepackage{xcolor}
\definecolor{linkColor}{RGB}{0,100,150}
\usepackage[colorlinks=true, allcolors=linkColor,pdfborder={0 0 0},pdfencoding = auto]{hyperref}

\usepackage{booktabs}
\usepackage{tabularx}

\usepackage{array}
\newcolumntype{L}[1]{>{\raggedright\let\newline\\\arraybackslash\hspace{0pt}}p{#1}}

\usepackage{tcolorbox}

\usepackage[sectionbib]{bibunits}
\usepackage{natbib}
\defaultbibliographystyle{pnas-new}
\defaultbibliography{GBE.bib}

\usepackage{hyphenat}

\usepackage{xcolor}
\definecolor{linkColor}{RGB}{0,70,120}
\definecolor{darkgreen}{RGB}{0,128,0}
\definecolor{darkgray}{RGB}{90,90,90}

\newcommand\Tvec{\mathbf{T}}
\newcommand\rvec{\mathbf{r}}

\newcommand\tvec{\mathbf{t}}
\newcommand\evec{\mathbf{e}}

\usepackage{bm}

\usepackage[normalem]{ulem}

\begin{document}

\title{A Geometric-tension-dynamics Model of Epithelial Convergent Extension}

\author{Nikolas H.\ Claussen}
\affiliation{Department of Physics, University of California Santa Barbara, Santa Barbara, California 93106, USA}
\author{Fridtjof Brauns}
\email{fbrauns@kitp.ucsb.edu}
\affiliation{Kavli Institute for Theoretical Physics, University of California Santa Barbara,
Santa Barbara, California 93106, USA}
\author{Boris I.\ Shraiman}
\email{shraiman@kitp.ucsb.edu}
\affiliation{Department of Physics, University of California Santa Barbara, Santa Barbara, California 93106, USA}
\affiliation{Kavli Institute for Theoretical Physics, University of California Santa Barbara,
Santa Barbara, California 93106, USA}

\begin{abstract}
Convergent extension of epithelial tissue is a key motif of animal morphogenesis. On a coarse scale, cell motion resembles laminar fluid flow; yet in contrast to a fluid, epithelial cells adhere to each other and maintain the tissue layer under actively generated internal tension.
To resolve this apparent paradox, we formulate a model in which tissue flow in the tension-dominated regime occurs through adiabatic remodeling of force balance in the network of adherens junctions.
We propose that the slow dynamics within the manifold of force-balanced configurations is driven by positive feedback on myosin-generated cytoskeletal tension.
Shifting force balance within a tension network causes active cell rearrangements (T1 transitions) resulting in net tissue deformation oriented by initial tension anisotropy.
Strikingly, we find that the total extent of tissue deformation depends on the initial cellular packing order.
T1s degrade this order so that tissue flow is self-limiting. We explain these findings by showing that coordination of T1s depends on coherence in local tension configurations, quantified by a geometric order parameter in tension space.
Our model reproduces the salient tissue- and cell-scale features of germ band elongation during \textit{Drosophila} gastrulation, in particular the slowdown of tissue flow after approximately twofold elongation concomitant with a loss of order in tension configurations.
This suggests local cell geometry contains morphogenetic information and yields experimentally testable predictions.
Defining biologically controlled active tension dynamics on the manifold of force-balanced states may provide a general approach to the description of morphogenetic flow.
\end{abstract}

\begingroup
\renewcommand\thefootnote{}\footnote{NHC contributed equally to this work with FB. NHC carried out the numerical simulations. All authors contributed to conceptualization, analysis, and writing of the manuscript}
\addtocounter{footnote}{-1}%
\endgroup

\maketitle





Shape changes of epithelia during animal development involve major cell rearrangements, often manifested as a ``convergent extension'' of cell sheets (CE).
On the coarse scale, CE resembles the laminar shear flow of an incompressible fluid in the vicinity of a hyperbolic fixed point (see Fig.~\ref{fig:intro}A).
Indeed, previous work has combined hydrodynamic equations for the mesoscale cell velocity field with active stress fields to model morphogenetic tissue flow \cite{Oster.etal1983,Streichan.etal2018,Saadaoui.etal2020,Ioratim-Uba.etal2023}.
Yet in contrast to a fluid, epithelia are under internally generated tension -- as revealed by laser ablation \cite{Kong.etal2019} -- and, like solids, maintain their shape against external forces.
Tissue flow is achieved through local cell intercalation (T1 neighbor exchange processes, see Fig.~\ref{fig:intro}B) driven by the concerted mechanical activity of individual cells. 
Cells generate forces via actomyosin contractility in the cortical cytoskeleton at the adherens junctions between cells (Fig.~\ref{fig:intro}C). 
Moreover, the adherens junctions can remodel through the turnover of their constituent molecules: interfaces in the cell array can change their length and tension independently.
This behavior is fundamentally different from (Hookean) springs, where tension and length are related by a constitutive relationship.
Instead, one can imagine cellular interfaces as ``microscopic muscles'' which are actuated by the recruitment and release of myosin motors. 

Vertex models generally describe epithelial tissue as a polygonal tiling of cells where the vertex positions are the dynamical variables \cite{Weliky.Oster1990,Farhadifar.etal2007}.
The forces that drive the vertex motion are commonly derived from a passive area and perimeter elasticity supplemented with additional active tensions \cite{Collinet.etal2015, Duclut.etal2022,Sknepnek.etal2023}.
However, the muscle metaphor for cellular interfaces suggests that \emph{active tension} is central to the mechanical network underlying an epithelial tissue (Fig.~\ref{fig:intro}D). This network rapidly equilibrates to a force-balanced state \cite{Bonnet.etal2012,Kong.etal2019,Noll.etal2020}, stabilized by mechanical feedback loops \cite{Noll.etal2017, Gustafson.etal2022}.
In such an active tension network, passive (bulk) elasticity plays a subdominant role.
The need for stabilizing feedback loops arises because active tensions are untethered from interface lengths. 
Indeed, on an abstract level, these feedback loops are not unlike the regulatory mechanisms that control and stabilize skeletal musculature \cite{Byrne1997}.

Here, we propose that tissue flow can be understood in terms of adiabatic (quasi-static) remodeling of internal active force balance.
Force balance in the cortical tension network defines a manifold of cellular tiling geometries on which tissue deformation unfolds.
We propose that dynamics in the force-balance manifold is driven by positive feedback on the cortical tensions.
This view is supported by analysis of high-quality live imaging data \cite{Stern.etal2022} from \textit{Drosophila} gastrulation presented in the companion paper \cite{Brauns.etal2024elife}. 
Specifically, a geometric active--passive decomposition as well as mutant analysis were used to show that tissue flow is driven by internally generated tension dynamics, rather than external forces.
Tension inference has provided evidence for the role of positive tension feedback during active T1 events. Numerical simulations of cell quartets show that such a feedback mechanism is sufficient to drive the T1 process.
However, the key question of coordination of T1s across the tissue -- required to drive coherent tissue flow -- has remained unanswered.
To address this question, we develop a model of tissue mechanics in the tension-dominated regime and demonstrate via numerical simulations how positive feedback drives CE.
We show that order of the cell packing is necessary for coordinating T1 processes, and hence efficient CE.
T1s destroy this order such that the extent of tissue flow is self-limiting.
Thereby, our model reproduces the experimentally observed elongation of the germband where the arrest of flow is concomitant with a transition from an ordered to a disordered cell packing \cite{Brauns.etal2024elife}. 

\section*{Methods}

\subsection*{A minimal model based on force balance and cell geometry}

\begin{figure*}[t]
    \centering
    \includegraphics{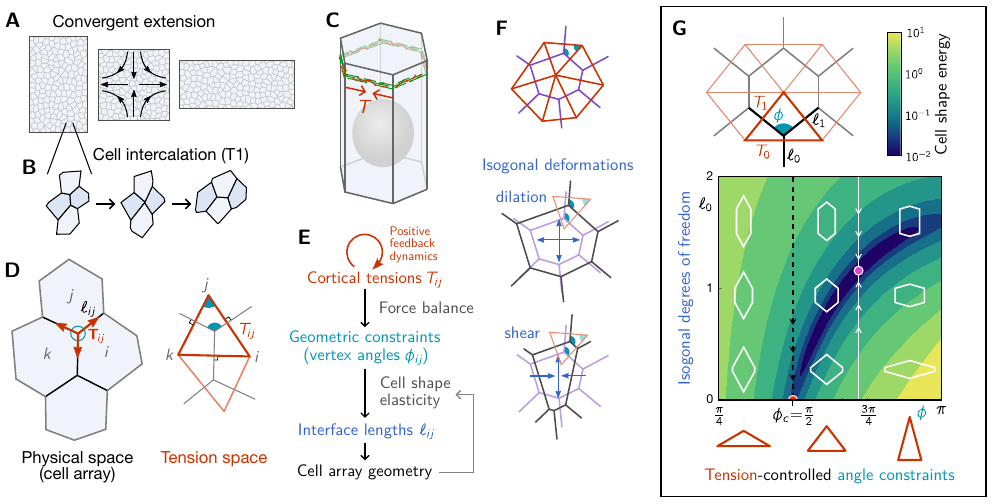}
    \caption{
    \textbf{A model for tissue mechanics dominated by cortical tension.}
    \textbf{A}~Convergent extension (CE) of epithelial tissue by cell intercalations (T1 processes, \textbf{B}).
    \textbf{C}~A single cell from a columnar epithelium with the actomyosin cortex at the adherens junction belt generating tension $T$ along cell-cell interfaces. Gray sphere represents the nucleus.
    \textbf{D}~In force balance, the tensile forces $\mathbf{T}_{ij}$ at each vertex must sum to zero, implying that they form a triangle. The angles in the tension triangulation are complementary to the angles in the cell array, thus linking tension space to physical space.
    Force balance implies that the tensile force vectors at each vertex form a triangle.
    \textbf{E}~Schematic outline of our quasi-static tissue model with mechanics dominated by actively regulated cortical tensions (see text for details).
    \textbf{F}~Illustration of isogonal deformations of a reference geometry (purple) that is dual to the tension triangulation (red).
    \textbf{G}~Implementation of the model for a symmetric, regular cell array, characterized by one angle $\phi$, determined by the tensions $T_0, T_1$, and two lengths, $\ell_0$ and $\ell_1$, parametrizing the soft isogonal modes. The contour plot shows the cell shape energy $E_\mathcal{C}$ in the incompressible limit where $\ell_1$ is determined uniquely by $\ell_0$ and $\phi$. 
    Relaxation of the sub-dominant cell shape energy $E_\mathcal{C}$ is constrained to the isogonal subspace (white line) determined by tension force balance. For the critical tension ratio $T_0/T_1 = \sqrt{2}$ (corresponding to $\phi_c = \pi/2$, black dashed line), the interface length minimizing $E_\mathcal{C}$ vanishes (red half-disk). See Movie~1 for an animated version.
    }
    \label{fig:intro}
\end{figure*}

Our model is based on two assumptions: (a) on morphogenetic timescales, the forces in the epithelium are approximately balanced, and (b) active cortical tensions (generated by contractile actomyosin along the adherens junctions Fig.~\ref{fig:intro}C) dominate over all other sources of stress.
In particular, we assume that adhesion forces between the epithelial layer and its substrate (the fluid yolk and perivitelline fluid \cite{Munster.etal2019} for the \textit{Drosophila} embryo) are negligible. Hence, all forces must be balanced within the transcellular network of cellular junctions.
We model the tissue in the framework of vertex models (see e.g.\ \cite{Farhadifar.etal2007}) as a polygonal tiling of the plane with tri-cellular vertices $\mathbf{r}_{ijk}$, where each polygon represents a cell $i$ (see Fig~\ref{fig:intro}D).
We write the elastic energy differential of this network as
\begin{equation} \label{eq:total-energy}
     dE(\{\mathbf{r}_{ijk}\} | \{T_{ij} \}) = \sum_{ij} T_{ij} d \ell_{ij} - p \sum_i d A_i + \varepsilon \sum_i d E_\mathcal{C}(S_i),
\end{equation}
where $\varepsilon$ is a small parameter that separates the dominant scale of active tension and subdominant passive mechanical contributions from bulk and shear elasticity of the cell interior.
$E_\mathcal{C}(S_i)$ accounts for the passive elasticity of the cells and will be specified below;
$\ell_{ij} = ||\mathbf{r}_{ij}||$ and $T_{ij}$ are the length of and tension along the interface between adjacent cells $i$ and $j$.
Importantly, in contrast to the standard vertex model where edge tension is defined by a constitutive relation corresponding to a passive Hookean perimeter spring, we take cortical tensions to be controlled independently of the interface lengths.
The tension dynamics is described in the next section.
The second term in Eq.~[\ref{eq:total-energy}] accounts for the effective in-plane pressure $p$ of the cells that, by maintaining the total surface area (sum over cell areas $A_i$), ensures that the tissue as a whole does not collapse.
In the \emph{Drosophila} embryo, the closed epithelial sheet encloses the yolk, which is under pressure, balancing the cortical tensions. Other epithelia are kept under tension via traction forces at the boundary \cite{Kunz.etal2023}.
We assume that pressure differences between cells are small and therefore absorb them into $E_\mathcal{C}(S_i)$. 
In experimental data, this can be verified by inspecting the curvature of cell-cell interfaces:
Due to the Laplace law, significant pressure differences would manifest through curved cell-cell interfaces which are not observed in the early \textit{Drosophila} embryo prior to the onset of cell divisions \cite{Brauns.etal2024elife,Stern.etal2022,Farrell.etal2017,Noll.etal2020}.

As noted, we assume a separation of scales between the timescale on which the elastic energy relaxes and the timescale on which the tissue deforms macroscopically. 
In terms of relaxational dynamics $\gamma \partial_t \mathbf{r}_{ijk} = -\frac{\partial  E}{\partial \mathbf{r}_{ijk}}$, we consider a relaxation rate -- set by the coefficient of friction $\gamma$ -- much faster than all other timescales in the system. Quasi-static force balance implies
\begin{align}
    \frac{\partial  E}{\partial \mathbf{r}_{ijk}} = 0.
\end{align}
Solving this equation to zeroth order in $\varepsilon$ yields a force-balance constraint at each vertex: the tension vectors $\mathbf{T}_{ij} = T_{ij} \mathbf{e}_{ij}$ at each vertex must sum to zero and hence form a triangle as illustrated in \ref{fig:intro}D. Since neighboring vertices share the interface that connects them, the corresponding tension triangles share an edge. Therefore, all tension triangles have to fit together: they form a triangulation that is dual to the cell tiling \cite{Noll.etal2017,Jensen.etal2020,Maxwell1864}.
This tension triangulation is a geometric manifestation of global force balance in the tissue, where angles at real-space vertices are complementary to the corresponding angles in the tension triangle (Fig.~\ref{fig:intro}D).

Importantly, fixing the angles at vertices \emph{does not} fully determine the cell tessellation, i.e.\ the $\mathbf{r}_{ijk}$: once can change the interface lengths $\ell_{ij}$ while preserving all angles.
The resulting \emph{isogonal soft modes}\footnote{A degree-of-freedom count shows there is one isogonal degree of freedom per cell \cite{Noll.etal2017}.
Therefore, the isogonal modes can be parametrized by an ``isogonal function'' that takes a scalar value in each cell. The isogonal displacement of a vertex is defined in terms of the values of this isogonal function in the three adjacent cells (see Eq.~[\ref{eq:isogonal_basis_vecs}] in the SI Appendix). Note that isogonal modes are only soft if they don't deform the tissue boundary.}
account for interface length changes under constant tension \cite{Noll.etal2017}, which is possible thanks to the turnover of cytoskeletal elements.

The isogonal modes can dilate and shear cells (Fig.~\ref{fig:intro}F).
They are soft modes of the leading order elastic energy and thus force us to take into account subleading contributions to arrive at a complete model of tissue mechanics.
Cells resist shape distortions due to rigid cell-internal structures such as microtubules, the nucleus \cite{Grosser.etal2021,Kim.etal2022nuclear},
and intermediate filaments \cite{Pensalfini.etal2023}.
To account for this passive cell elasticity, we propose an energy
\begin{equation}
   E_\mathcal{C}(S) = \lambda [\mathrm{Tr}(S-S_0)]^2 + \mu \mathrm{Tr}[(S-S_0)^2], 
   \label{eq:cell_shape_energy}
\end{equation}
in terms of the cell shape tensor 
\begin{equation}
    S_{i} = \sum_{k \in \mathcal{N}_i} \frac{ \mathbf{r}_{ik} \otimes \mathbf{r}_{ik}}{\ell_{ik}},
\end{equation}
where $\mathcal{N}_i$ is the set neighbors of cell $i$.
This shape energy effectively models the cell interior as a homogeneous elastic material \cite{Wohlrab.etal2024}.
(In SI Sec.~\ref{SI:simulation_methods}, we compare this elastic energy with the often-used ``area--perimeter'' elastic energy which we find produces qualitatively different behavior incompatible with experimental observations.)
The shape tensor is defined to be invariant under subdivision of interfaces.
The reference tensor $S_0$ controls the target cell shape and is given by $S_0 = 3 \ell_0 \, \mathbb{I}$ for an isotropic hexagonal cell with side length $\ell_0$. 
We relate cell and tissue elasticity by analyzing the energy spectrum of isogonal modes for a fixed tension triangulation (see SI Sec.~\ref{SI:iso-shear}). 
The isogonal modes with the lowest energy correspond to large-scale shears and thus provide a linear relationship between the cell and tissue shear moduli (see SI Fig.~\ref{SI-fig:isogonal-shear})\footnote{As explained in the SI, this analysis only applies for deformations in the bulk of the cell array that leaving the boundary shape unchanged.}.

For sufficiently small values of the scale-separation parameter $\varepsilon$, minimization of the elastic energy Eq.~[\ref{eq:total-energy}] can be performed in two separate steps: First, force balance of the dominant cortical tensions $T_{ij}$ fixes the angles at vertices thus setting geometric constraints. Second, the sub-leading term $\sum_i E_\mathcal{C}(S_i)$ is minimized to fix the remaining isogonal soft modes.
Importantly, in this limiting case, the value of $\varepsilon$ is immaterial as long as it is small enough (see SI Sec.~\ref{SI:additional-simulations}).
Figure~\ref{fig:intro}G and Movie~1 illustrate the minimization of the cell shape energy under the angle constraints imposed by junctional force balance for the minimal setting of a perfectly symmetric cell array. The geometry is characterized by two interface lengths, $\ell_0$, $\ell_1$, and single angle $\phi$. 
Incompressibility fixes $\ell_1$ as a function of $\ell_0$ and $\phi$. 
We can then plot the cell shape energy $E_\mathcal{C}$ in a two-dimensional $\phi, \ell_0$ energy landscape.
Force balance of the tensions $T_0$, $T_1$ constrains the angle $\phi = 2\arccos(T_0/2T_1))$ and relaxation of $E_\mathcal{C}$ takes place in the isogonal subspace parameterized by $\ell_0$ (vertical white line in Fig.~\ref{fig:intro}G), thus fully determining the cell geometry (purple dot).
Changing the tensions $T_0, T_1$ shifts the angle constraint and therefore forces the cell shape energy to relax to a new cell geometry.
By changing constraints, the dynamics of the tension configuration drives tissue flow. 
When the tensions reach the critical ratio $T_0/T_1 = \sqrt{2}$ such that $\phi = \pi/2$, the length of the vertical interface, $\ell_0$, vanishes, causing a T1 transition as discussed in the companion paper \cite{Brauns.etal2024elife}.

\subsection*{Positive feedback and adiabatic dynamics}

On the timescale of morphogenetic flow, tensions change due to the recruitment and release of molecular motors, driving the remodeling of the force balance geometry encoded in the tension triangulation. 
To complete the model, we need to specify the dynamics that governs the tensions on this slow timescale.

Based on previous experiments \cite{Fernandez-Gonzalez.etal2009} and models \cite{Odell.etal1981,Sknepnek.etal2023}, we propose a positive feedback mechanism where tension leads to further recruitment of myosin motors and thus further increase in tension. 
This self-amplifying recruitment is limited by the competition for a limited pool of myosin within each cell. 
To mimic this effect in a computationally simple way, we constrain tension dynamics to conserve the perimeter of each tension triangle, i.e.\ the sum of tensions at each vertex $(ijk)$ (see SI Sec.~\ref{SI:additional-simulations} for different local conservation laws).
For an individual triangle with tensions $\tilde{T}_1, \tilde{T}_2, \tilde{T}_3$, we consider the dynamics
\begin{equation} \label{eq:tension-dynamics}
    \tau_\mathrm{T}^{} \partial_t \tilde{T}_{\alpha} = \tilde{T}_{\alpha}^{n} - \frac{1}{3} \sum_{\beta=1}^3 \tilde{T}_{\beta}^n, \quad \text{with} \quad \alpha = \{1, 2, 3\},
\end{equation}
where $n$ is an exponent that determines the nonlinearity of the feedback.
Note that each cell-cell interface is composed of two actomyosin cortices on its two sides and only one of the two is part of each local pool (see SI Sec.~\ref{SI:simulation_methods} for details).
This feedback mechanism has a ``winner-takes-all'' character, where the longest edge in the tension triangle always outgrows the other two.
In our model framework, we can consider a variety of possible local tension dynamics. Below we will also investigate a form of positive tension feedback that saturates, and identify the qualitative features of local tension dynamics key to the tissue dynamics.

Force balance requires that all tension triangles fit together to form a flat triangulation \cite{Noll.etal2017}.
The triangulation is parameterized by a set of 2D tension vertex positions $\mathbf{t}_i$, so that the tension on edge $(ij)$ is given by $T_{ij}=||\mathbf{t}_i-\mathbf{t}_j||$.
In each iteration of the simulation, the tension vertices $\mathbf{t}_i$ are determined by fitting the balanced tensions $T_{ij}$ to the intrinsic tensions $\tilde{T}_{ij}$ using a least squares method.
In addition, the intrinsic tensions $\tilde{T}_{ij}$ relax to the balanced tensions $T_{ij}$ with a rate $\tau_\mathrm{balance}^{-1} \ll \tau_\mathrm{T}^{-1}$ (see SI Sec.~\ref{SI:simulation_methods} for details. All quantifications presented here refer to the flat tensions $T_{ij}$.).
This ``balancing'' of the tension triangulation effectively accounts for small pressure differentials and additional feedback mechanisms (such as the strain rate feedback \cite{Noll.etal2017, Gustafson.etal2022}) which maintain the tension network in a state of force balance. In particular, this ensures that cortical tensions do not lead to a build-up of pressure differentials.

The above dynamics is autonomous in tension space until an edge in the cell tessellation reaches length zero. At this point, a cell neighbor exchange (T1 transition) occurs, corresponding to an edge flip in the tension triangulation. After this topological modification, the tension dynamics continues autonomously again until the next T1 event. 
To determine the active tension (i.e.\ myosin level) on the new interface formed during the cell neighbor exchange, we assume continuity of myosin concentration at vertices as described in the companion paper \cite{Brauns.etal2024elife} and in SI Sec.~\ref{SI:simulation_methods}. The active tension is not sufficient to balance the total tension on the new interface, such that passive elements of the cortex (e.g.\ crosslinkers) are transiently loaded. The resulting passive tension relaxes due to remodeling with timescale $\tau_\mathrm{p}$ (see SI Sec.~\ref{SI:simulation_methods}) \cite{Clement.etal2017}. This relaxation causes the elongation of the new interface, transiently counteracting positive tension feedback and thereby prevents the new interface from immediately re-collapsing after a T1.

This concludes the description of the computational model.
A brief overview of the parameters and their effects is given in SI Table~\ref{table:simulation_defaults} and Fig.~\ref{SI-fig:parameter_scan}. Fig.~\ref{SI-fig:simulation_flow_chart} provides a flow chart of the simulation algorithm.

\begin{figure*}[t]
    \centering
    \includegraphics[width=\textwidth]{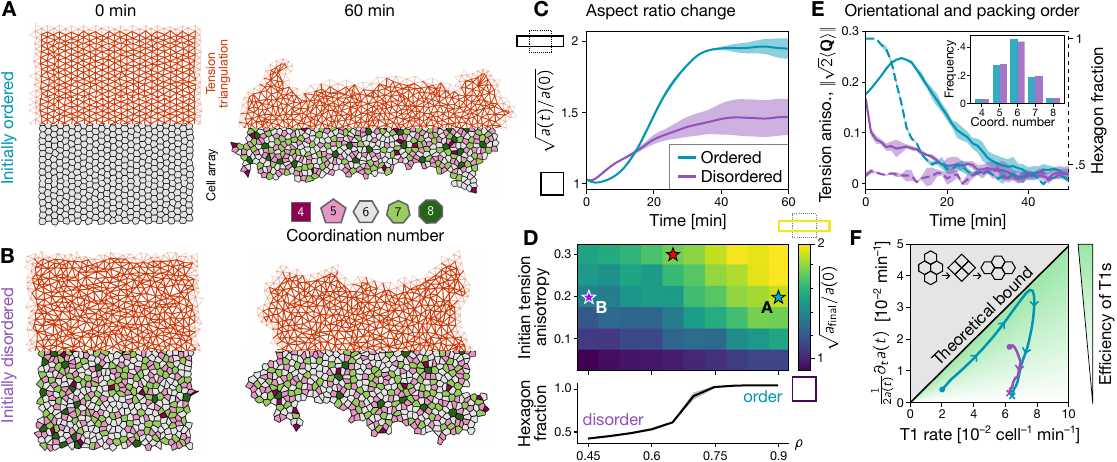}
    \caption{
    \textbf{Extent of tissue flow depends on initial cell-scale order and tension anisotropy.}
    \textbf{A, B}~Simulation snapshots of tissue patches with free boundaries starting from an (irregular) hexagonal cell array (\textbf{A}) and a disordered cell array (\textbf{B}) generated from a random Voronoi tessellation. As small initial tension anisotropy orients convergent-extension flow. Snapshots show the top half of the tension triangulation and the bottom half of the corresponding cell array.
    \textbf{C}~CE (measured by the tissue aspect ratio $a(t)$) is slower and ceases at a smaller net deformation for a disordered initial condition (cyan line) compared to the initially ordered case (purple line).
    Shaded bands indicate standard deviation over $N=3$ simulation runs with $N_\mathrm{cells}\approx 10^3$ cells each.
    \textbf{D}~The total aspect ratio change $a_\mathrm{final}/a(0)$ increases as a function of cell packing order (hexagon fraction controlled by hard disk packing fraction $\rho$ used to generate the Voronoi seed points, see bottom plot) and initial tension anisotropy (controlled by a shear $s$ applied to the tension triangulation).
    The red star indicates the approximate state of the \textit{Drosophila} germ band at the onset of germ band extension (cf.\ Figs.~\ref{SI-fig:hexagonal-order}B, \ref{SI-fig:local-vs-regional-inference}), corresponding to a $1.85\times$ elongation.
    \textbf{E}~Decreasing hexagon fraction (solid lines) and global tension anisotropy (dashed lines) indicate the decay of order in the cell arrays. Notably, in the initially ordered cell array, tension anisotropy transiently increases due to positive tension feedback (solid purple line). At late times both simulations converge to zero global tension anisotropy and identical coordination number statistics (inset).
    \textbf{F}~Plotting the rate of tissue CE against the T1 rate provides a measure for the efficiency of T1 transitions. For the initially ordered tissue, the efficiency of T1s starts out near the theoretical optimum ($\tfrac{1}{2}\log{3}$ elongation for one T1 per cell as illustrated in the inset cartoon) but drops to zero as the tissue becomes disordered.}
    \label{fig:tissue-model-disorder}
\end{figure*}

\section*{Results}
 
\subsection*{Cell packing order facilitates self-organized convergent-extension flow}

In the companion paper \cite{Brauns.etal2024elife}, we have shown that positive tension feedback can drive active T1 transitions in a regular lattice of cells with an initial anisotropy of tension.
Any real tissue will exhibit some degree of irregularity. Therefore, investigating the effect of this disorder is key to understanding CE on the tissue scale.
To this end, we perform simulations of freely suspended irregular cell arrays. All parameters are set to the same values as in the companion paper, where they were calibrated to fit the tension and interface length dynamics of active T1s during \textit{Drosophila} gastrulation (see Fig.~\ref{SI-fig:T1-dynamics}).

We generated initial tension triangulations from random hard disk packings at different packing fractions $\rho$ \cite{Bernard.etal2009}. 
At low packing fraction, the hard disk process generates highly irregular triangulations (see Fig.~\ref{fig:tissue-model-disorder}B) while at sufficiently high packing fraction $\rho \gtrsim 0.72$, the disks adopt a crystalline packing such that the fraction of cells with six neighbors $p_6 \approx 1$ (Fig.~\ref{fig:tissue-model-disorder}A).
To introduce a specified initial tension anisotropy, the triangulation is sheared with magnitude $s$ (displacing vertices by $\mathbf{t}_i \mapsto \mathrm{diag}(\sqrt{1 - s}, 1/\sqrt{1 + s}) \, \mathbf{t}_i$).

To quantify tension anisotropy, we define the tensor $Q=\tfrac{2}{3}\sum_{\alpha=1}^{3} \mathbf{T}_\alpha\otimes \mathbf{T}_\alpha$ for each triangle directly from the tension geometry. Averaging its deviatoric part $\tilde{Q}=Q - \frac{1}{2} \mathrm{Tr}\,Q$ over the cell array, provides a measure of global tension anisotropy $||\langle \sqrt{2}\tilde{Q} \rangle|| \in [0,1]$.
Starting with a slightly perturbed hexagonal cell packing and a small initial tension anisotropy, the tissue patch undergoes CE, elongating perpendicular to the initial orientation of global tension anisotropy (Fig.~\ref{fig:tissue-model-disorder}A).

The tissue flow is driven by self-organized cell rearrangements (active T1 transitions) whose rate rapidly increases, reaching a maximum, and then decreases to a lower, but non-zero, value (Fig.~\ref{fig:tissue-model-disorder}F).
Large-scale tissue deformation stalls after approximately 2-fold CE (as measured by the square root of the aspect ratio $a=\mathrm{width}/\mathrm{height}$, Fig.~\ref{fig:tissue-model-disorder}C) while cells continue rearranging.
T1s at this stage are no longer coherently oriented and therefore do not contribute to net tissue deformation (Fig.~\ref{fig:tissue-model-disorder}F). 

As cells rearrange, the tissue becomes increasingly disordered, as indicated by the loss of global tension anisotropy and the decreasing fraction of cells with six neighbors, $p_6$ (Fig.~\ref{fig:tissue-model-disorder}E). The peak of tension anisotropy in the ordered initial condition coincides with the onset of T1 transitions and the decay of $p_6$ (Fig.~\ref{SI-fig:parameter_scan}).

By contrast, initializing the simulation with a low level of order in the initial cell packing, but identical tension anisotropy, results in slower convergent-extension flow and arrest of flow at a smaller amount of total tissue-scale deformation (Figs.~\ref{fig:tissue-model-disorder}B and \ref{fig:tissue-model-disorder}C and Movie~2).
Notably, tension anisotropy rapidly vanishes without the transient increase observed in the simulation starting with a more ordered cell packing (Fig.~\ref{fig:tissue-model-disorder}E).
While the early dynamics depends sensitively on the initial condition, we find rapid convergence toward a common disordered steady state.

The heatmap in Fig.~\ref{fig:tissue-model-disorder}D shows the dependence of CE on the initial configuration as controlled by $\rho$ and $s$. 
The total extent of CE, quantified by the net change in aspect ratio $\sqrt{a_\mathrm{final}/a_\mathrm{initial}}$ increases as a function of the initial order and the magnitude of tension anisotropy.
The degradation of order through cell rearrangements means that the system dynamically traverses the phase space spanned by order $p_6$ and anisotropy $||\langle \sqrt{2}\tilde{Q} \rangle||$ and the remaining extension is predicted by the instantaneous value of these two quantities (see SI Fig.~\ref{SI-fig:disorder}).

We find that all simulations converge to a disordered state where T1s are incoherent and tissue flow stalls.
This naturally explains key aspects of germ-band extension in the \textit{Drosophila} embryo, in particular, the transition from the fast to the slow phase of germ-band extension \cite{Brauns.etal2024elife}, concomitant with an increase in cell-scale disorder, approaching a maximally disordered state \cite{Irvine.Wieschaus1994,Brauns.etal2024elife} (see experimental data in Fig.~\ref{SI-fig:hexagonal-order}).
The self-limiting character of CE driven by positive tension feedback is robust across variations of the tension dynamics model (see SI Fig.~\ref{SI-fig:parameter_scan}, SI Sec.~\ref{SI:additional-simulations}, and SI Table~\ref{table:simulation_defaults}) and a similar phenomenon was observed in a recent model by Sknepnek \textit{et al.} \cite{Sknepnek.etal2023}. 

\subsection*{Order in local tension configurations}

\begin{figure}[t]
    \centering
    \includegraphics{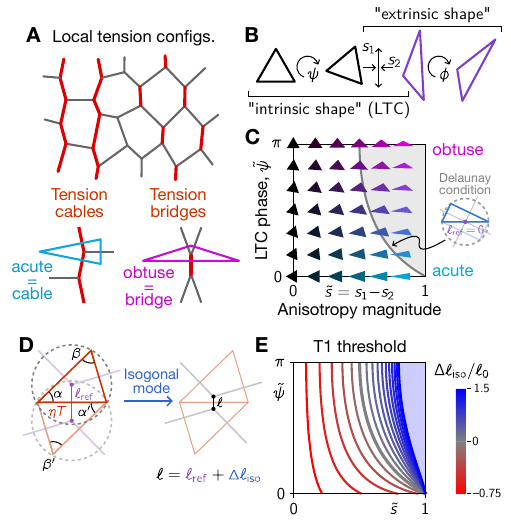}
    \caption{
    \textbf{Triangle shape space characterizes the local tension motifs that underlie cell rearrangements.}
    \textbf{A}~Local tension configurations, ``cables'' and ``bridges'', can be quantified in terms of the tension triangle shapes.
    \textbf{B}~Shape decomposition of a triangle into a sequence of three transformations acting on an equilateral reference triangle. The angle $\psi$ and the stretch factors $s_{1,2}$ determine the intrinsic shape of the triangle while the angle $\phi$ determines its orientation in space.
    \textbf{C}~Intrinsic shape space of triangles parameterized by anisotropy (LTC magnitude) $\tilde{s}$ the ``LTC phase'' $\tilde{\psi}$ that distinguishes obtuse and acute triangles.
    The gray line indicates the Delaunay condition for a pair of identical triangles. Along this line, the circumcenters of the two triangles coincide, corresponding to a fourfold vertex in the Voronoi tessellation.
    \textbf{D}~Circumcircle construction of the Voronoi edge length $\ell_\mathrm{ref}$ (purple) from a pair of adjacent (tension) triangles (red). Circumcircles are indicated by gray dashed lines. The actual physical length $\ell$ is the sum of the Voronoi reference length $\ell_\mathrm{ref}$, and a contribution from isogonal strain $\Delta \ell_\mathrm{iso}$. In the illustration, $\Delta \ell_\mathrm{iso}/\ell_0$ is negative.
    \textbf{E}~T1 threshold as a function of the isogonal strain $\Delta \ell_\mathrm{iso}/\ell_0$. Positive isogonal strain shifts the threshold to higher tension anisotropy.
    }
    \label{fig:tension-config-space}
\end{figure}

So far, we have focused on the role of tension anisotropy and initial topological order in the cell packing. 
In the \textit{Drosophila} germ band, we additionally observed a more subtle form of geometric order -- a particular pattern of alternating high and low tensions \cite{Brauns.etal2024elife} -- that arises dynamically before the onset of cell rearrangements.
The elementary motifs of a cell-scale tension pattern are the tension triangles at individual vertices. Acute triangles correspond to tension cables (adjacent high-tension interfaces) while obtuse triangles -- to which we refer as tension ``bridges'' -- are the elementary motif of an alternating pattern of high and low tensions (Fig.~\ref{fig:tension-config-space}A).

To quantify the relative abundance of these motifs and compare our simulations to experimental data we define a local tension configuration (LTC) order parameter that measures how anisotropic and how acute vs obtuse a given tension triangle is.
To construct this order parameter, the three tension vectors $\mathbf{T}_\alpha, \alpha = 1,2,3$ that form the tension triangle are first ordered by increasing length, i.e.\ $T_1\leq T_2\leq T_3$,
and then combined into a 2$\times$2 matrix
\begin{equation}
    \mathfrak{T} =
    \begin{pmatrix}
        \bm{\tau}_1 \\
        \bm{\tau}_2
    \end{pmatrix} =
    \frac{1}{\sqrt{2} \, \mathcal{N}} \begin{pmatrix}
        T_1^x - T_2^x & T_1^y - T_2^y \\
        \sqrt{3} \, T_3^x & \sqrt{3} \, T_3^y 
    \end{pmatrix}.
    \label{eq:barycentric_matrix}
\end{equation}
The normalization factor $\mathcal{N}$ ensures $||\mathfrak{T}||^2 = \mathrm{Tr}[\mathfrak{T}\mathfrak{T}^T] = 1$, fixing the arbitrary overall tension scale.
$\mathfrak{T}$ is not a symmetric matrix and its indices belong to different spaces: the first labels the barycentric component and the second the Cartesian coordinate.
We now carry out a singular value decomposition (SVD, geometrically illustrated in Fig.~\ref{fig:tension-config-space}B):\footnote{A similar decomposition was used in Ref.~\cite{Merkel.etal2017} to quantify tissue strain rates from a cell-centroid-based triangulation. However, the information contained in the ``LTC phase'' $\psi$ was not utilized there.}
\begin{equation} \label{eq:SVD}
    \mathfrak{T} = R(\psi) \cdot \begin{pmatrix}
    \sqrt{s_1} & 0 \\ 0 & \sqrt{s_2}
    \end{pmatrix} \cdot C \cdot R^{T}(\phi),
\end{equation}
where $R(\alpha)$ is the rotation matrix with angle $\alpha$, the singular values are ordered $s_1>s_2>0$ by convention, and $s_1 + s_2 = 1$ because we have normalized the tension vectors.
The reflection matrix $C=\mathrm{diag}(1, \pm 1)$ accounts for the chirality of the tension triangle (i.e.\ whether the edges go clockwise or counterclockwise if sorted by length).
The angle $\phi$ represents the orientation of tension anisotropy in physical space. Indeed, the triangle anisotropy tensor $Q$ defined above is given by $Q = \mathfrak{T}^T\mathfrak{T}$.
The \emph{intrinsic} rotation angle $\psi \in[0, \pi/6]$ controls whether the subsequent shear $\mathrm{diag} \, (\sqrt{s_1}, \sqrt{s_2})$ makes the triangle obtuse or acute. 
The intrinsic shape properties of the triangle are therefore parametrized by the magnitude of anisotropy $\tilde{s}:=(s_1 - s_2) \in[0,1]$ and the ``LTC phase'' $\tilde{\psi} = 6\psi\in[0,\pi]$ mapping out a two-dimensional shape space, which we refer to as LTC space (Fig.~\ref{fig:tension-config-space}C).

The SVD Eq.~[\ref{eq:SVD}] links two spaces with different symmetries,  the hexatic symmetry of the tension triangulation with the nematic symmetry of deviatoric stress in physical space. Thus, $\mathfrak{T}$ represents a \emph{hexanematic field}. 
In contrast, the hexanematic cross-correlation defined in \cite{Armengol-Collado.etal2024}, $\mathfrak{T}$ is purely local and does not depend on the magnitude of hexatic order.
Moreover, the nematic order represents the orientation of deviatoric stress, not cell elongation as in \cite{Armengol-Collado.etal2023,Armengol-Collado.etal2024}. This distinction is important because cell elongation is controlled by the isogonal modes and can therefore decouple from stress anisotropy.

\subsection*{A generalized Delaunay condition defines the locus of T1 events in LTC space}

The tight coupling between tension space and physical space allows us to define a condition for the occurrence of T1 transitions in LTC space.
This T1 threshold will allow us to quantify how tension dynamics causes active T1s by driving the local tension configurations towards the T1 threshold.
It also puts a constraint on the local tension configurations that we expect to observe.

Let us for a moment neglect the isogonal modes. From the tension triangulation we construct the corresponding Voronoi tessellation whose vertices are the circumcircle centers of the triangles as illustrated in Fig.~\ref{fig:tension-config-space}D. The edges of the Voronoi tessellation are orthogonal to those of the triangulation, which implies that it obeys the force balance constraints, and can be used as a reference for the family of cell arrays compatible with the tension triangulation. The length of a Voronoi edge corresponding to a pair of adjacent triangles is given by
\begin{equation} \label{eq:Voronoi-length}
    \ell_\mathrm{ref} = \frac{\sqrt{3} \, \ell_0 T}{2} (\cot{\beta} + \cot{\beta'}),
\end{equation}
where $T$ is the length of the shared triangle edge interface and $\ell_0$ fixes the length scale such that $\ell_\mathrm{ref}=T$ for equilateral tension triangles.
$\ell_\mathrm{ref}$ changes sign at $\beta + \beta' = \pi$, which gives the ``Delaunay condition'' $\beta + \beta' < \pi$.
In the absence of isogonal strain, a cell neighbor exchange (corresponding to an edge flip in the triangulation) must occur upon crossing this threshold. 
In Fig.~\ref{fig:tension-config-space}C, the gray line indicates this threshold for a pair of identical triangles (i.e.\ $\beta = \beta' = \pi/2$).
Notably, the threshold is at a much smaller anisotropy magnitude $\tilde{s}$ for tension cables (small $\tilde{\psi}$) than for bridges (large $\tilde{\psi}$), implying that tension cables are less efficient at driving intercalations than tension bridges.

How does the Delaunay condition generalize in the presence of isogonal strain?
The length of the central interface, $\ell$, can be decomposed as
\begin{equation}
    \ell = \ell_\text{ref} + \Delta \ell_\text{iso},
\end{equation}
where the isogonal contribution $\Delta \ell_\text{iso}$ accounts for isogonal modes while the (Voronoi) reference length is given by Eq.~[\ref{eq:Voronoi-length}].
Note that $\Delta \ell_\text{iso}$ is not an edge-autonomous quantity but depends on the isogonal mode (parameterized by the isogonal function) in the four cells surrounding the interface. In practice, $\Delta \ell_\text{iso}$ can be estimated from the average isogonal strain tensor in a local tissue patch \cite{Brauns.etal2024elife}. 
Now an interface collapses if the \emph{physical} length reaches zero: $\ell_\text{ref} + \Delta \ell_\text{iso} = 0$. This generalizes the Delaunay condition. Figure~\ref{fig:tension-config-space}E shows the shifted T1 threshold as a function of the isogonal strain $\Delta \ell_\mathrm{iso}/\ell_0$ (see SI for a mathematical expression).

\subsection*{Winner-takes-all feedback drives coherent T1s through formation of tension bridges}

\begin{figure}[tb]
    \centering
    \includegraphics{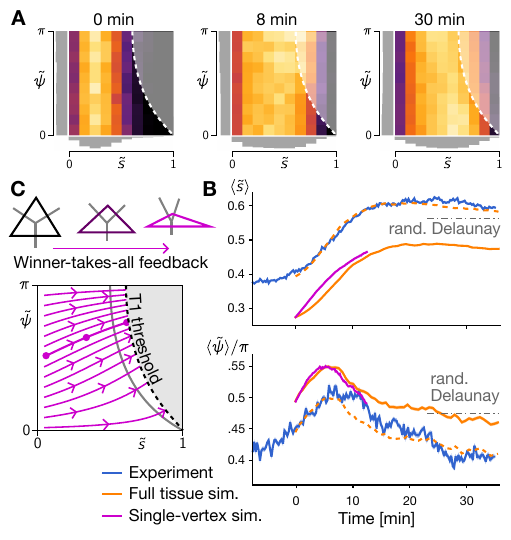}
    \caption{
    \textbf{Dynamics of LTC order in simulations and the \textit{Drosophila} germ band.}
    \textbf{A}~Heat maps showing the distribution of local tension configurations in a simulation with imposed isogonal strain matching the experimental observations.
    Data aggregated from $N=6$ simulation runs of $\sim10^3$ cells each.
    \textbf{B}~Driven by winner-takes-all feedback, the magnitude of tension anisotropy magnitude $\tilde{s}$ (top) and the tension bridge fraction (measured by the LTC phase $\tilde{\psi}$; bottom) increase. As the tissue becomes disordered due to cell rearrangements, the bridge fraction starts decreasing at ca.\ 10~min. Solid lines show the median of the LTC distributions where the phase $\tilde{\psi}$ is weighted with the magnitude $\tilde{s}$. The width of bands showing the standard error is comparable to the line width. Shifting the median from simulations by a constant offset (dashed orange lines) yields a quantitative match to the experimental data. 
    \textbf{C}~Shape dynamics of a single tension triangle driven by winner-takes-all feedback rapidly drives the tensions towards the T1 threshold with a slight bias towards tension bridges.
    }
    \label{fig:tension-config-data}
\end{figure}

The LTC order parameter and the T1 threshold in hand, we can quantify the dynamics of tensions in the simulations (see Fig.~\ref{fig:tension-config-data}A) and experiments (see companion paper Ref.~\cite{Brauns.etal2024elife} and Fig.~\ref{SI-fig:local-vs-regional-inference}).
Because isogonal strain shifts the T1 threshold (cf.\ Fig.~\ref{fig:tension-config-space}E), it will have a significant effect on the LTC order parameter distribution.
We imposed in our simulations the isogonal strain observed in the \textit{Drosophila} germ band \cite{Brauns.etal2024elife}, where invagination of the adjacent mesoderm tissue causes isogonal strain along the axis of tension anisotropy (see SI Sec.~\ref{SI:additional-simulations}).

LTC histograms show an increase in anisotropy and a transient bias towards tension bridges, before convergence to a steady state biased towards tension cables as the tissue becomes disordered (Fig.~\ref{fig:tension-config-data}A). 
Time traces of the median anisotropy $\tilde{s}$ and (weighted) median LTC phase $\tilde{\psi}$ show qualitative agreement with the experimental data from the \textit{Drosophila} germ band \cite{Brauns.etal2024elife} (Fig.~\ref{fig:tension-config-data}B). Quantitative agreement can be achieved by adding constant offsets (dashed lies in Fig.~\ref{fig:tension-config-data}B), which may be a consequence of noise in the experimental data (see SI-Fig. \ref{SI-fig:parameter_scan} for simulations incorporating Langevin noise). 

To understand the LTC dynamics, consider the shape dynamics of a single, isolated tension triangle governed by winner-takes-all feedback, Eq.~\ref{eq:tension-dynamics}].
Starting from a configuration with nearly equal tensions, the highest tension grows at the expense of the other two, driving the triangle towards an increasingly anisotropic and obtuse shape, as illustrated in Fig.~\ref{fig:tension-config-data}C. This LTC flow drives the tension configurations towards the T1 threshold and thereby causes cell rearrangements.
The single-triangle simulation successfully predicts early dynamics of the LTC distribution until the onset of cell rearrangements (Fig.~\ref{fig:tension-config-data}B).
The single-triangle picture also highlights the impact of isogonal strain; in simulations without imposed isogonal stretching, the T1--threshold is positioned so that bridges are rapidly eliminated by T1s (gray line in Fig.~\ref{fig:tension-config-data}C), and no transient bride bias is observed. We predict that this will occur in \textit{twist} or \textit{snail} mutant embryos where mesoderm invagination is abolished.

While positive tension feedback explains the emergence of tension bridges at the \emph{local} (single-triangle) level, it is not enough to produce an alternating pattern of tensions across cells.
For such a pattern, tension bridges must fit together coherently, i.e.\ their tension anisotropy is aligned across cells.
This requires that the coordination number of a majority of cells is 6, i.e.\ that most cells are hexagons.
This explains why some degree of hexagonal packing order is required to drive coherent T1s that underlie rapid CE.
Notably, however, long-range hexatic order is not necessary, as the local positive feedback on tensions is able to promote the local alignment of hexatic and nematic order in tension space, as manifest in the LTC phase dynamics.

In Fig.~\ref{fig:tissue-model-disorder}, we have seen that as cells rearrange, the cell array becomes disordered with the coordination number statistics approaching a random Voronoi tessellation. This suggests that the corresponding tension triangulations resemble random Delaunay triangulations.
To generate a family of such random triangulations, we use the same hard disk sampling method as above, controlled by the packing fraction $\rho$. 
We find that the triangle shape (LTC) statistics of a random Delaunay triangulation with $\rho \approx 0.2$ reproduces the late time statistics observed in simulations and in the \textit{Drosophila} germ  (see Fig.~\ref{SI-fig:LTC-Random-Delaunay}).
Notably, the late-time distribution exhibits a slight bias towards tension cables. The loss of tension bridges causes active T1s to become incompatible between adjacent cells, contributing to a slowdown of tissue extension found in tissue scale simulations and in the germ band \cite{Brauns.etal2024elife}.

Taken together, we find that the time course of LTC distribution agrees between the model and the experimental data. 
Next, we show how changing aspects of the model affects the LTC distribution, highlighting that the LTC parameter can be used to distinguish different tension dynamics based on \emph{statistical} signatures of cell-scale observations.

\subsection*{Saturating tension feedback causes tension cable formation and reduced convergence extension}

\begin{figure}[tb]
    \centering
    \includegraphics{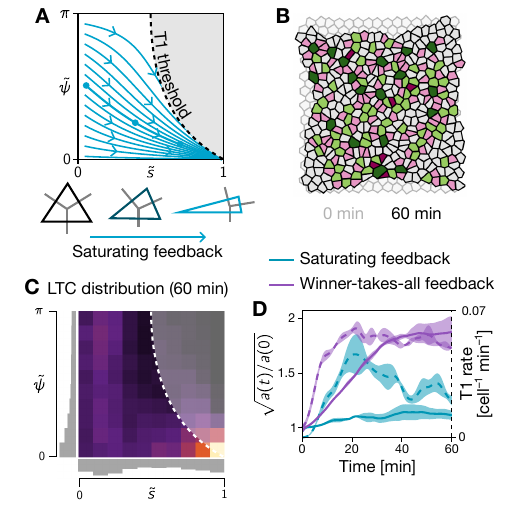}
    \caption{
    \textbf{Saturating tension feedback causes tension cable formation and, hence, fails to drive CE.}
    \textbf{A}~The flow in local tension configuration space induced by saturating positive feedback is inefficient at driving T1s.
    \textbf{B}~Saturating feedback generates only little CE (quantification in D). The initial configuration is shown semi-transparent in the background. Cell color indicates coordination number (cf.\ Fig.~\ref{fig:tissue-model-disorder}A).
    \textbf{C}~The late-time LTC distribution shows a strong cable bias and differs significantly from the random Delaunay distribution emerging in simulations with winner-takes-all feedback (cf.\ Fig.~\ref{fig:tension-config-data}D).
    \textbf{D}~Saturating feedback (teal), compared to winner-takes-all feedback (purple), yields very little aspect ratio change and a significantly reduced T1 rate.
    }
    \label{fig:saturating-feedback}
\end{figure}

The ``winner-takes-all'' local tension feedback mechanism, Eq.~[\ref{eq:tension-dynamics}], considered so far is efficient at driving T1s because it causes the formation of tension bridges as illustrated in Fig.~\ref{fig:tension-config-data}C.
In contrast, when positive feedback rapidly saturates, adjacent high tension interfaces no longer compete, leading to the formation of tension cables (Fig.~\ref{fig:saturating-feedback}A; see SI Sec.~\ref{SI:additional-simulations} for details). 
The trajectories in LTC space obtained from single-triangle simulations show that saturating feedback is less efficient at driving the local tension configuration towards the T1 threshold.
Indeed, tissue scale simulations with such feedback produce very little CE (see Figs.~\ref{fig:saturating-feedback}B and~\ref{fig:saturating-feedback}D and Movie~3).
The rate of T1 transitions is significantly reduced  (Fig.~\ref{fig:saturating-feedback}D), and in contrast to ``winner-takes-all'' feedback, a significant fraction of T1 transitions (approximately 20\%) are reversible, i.e.\ the newly formed edge rapidly re-collapses (see SI Sec.~\ref{SI:additional-simulations}).
Indeed, the T1 rate is transiently quite high but there is very little CE, suggesting that T1s along cables are inefficient at driving tissue deformation.
Saturating tension feedback might therefore explain the reversible T1s observed in certain \textit{Drosophila} mutants \cite{Bardet.etal2013}.
As predicted from single-triangle shape-space flow (Fig.~\ref{fig:saturating-feedback}A) the LTC distribution develops a significant bias towards tension cables as shown in Fig.~\ref{fig:saturating-feedback}C.
Such persistent tension cables are observed in \textit{Drosophila abl} mutants, suggesting that knockout of \emph{abl} might impair positive tension feedback (see Discussion).

The above findings show that the LTC order parameters capture important structural features on the cellular scale that strongly affect the dynamics and efficiency of T1s processes.
Flow in LTC parameter space obtained from single-triangle simulations serves as a simple tool to predict the cell scale behavior (efficiency of active T1s, emergence of tension cables vs bridges) for a given tension-feedback law.

\subsection*{Tension-triangulation model reproduces \textit{Drosophila} axis elongation in a simplified geometry}

\begin{figure*}[t]
    \centering
    \includegraphics{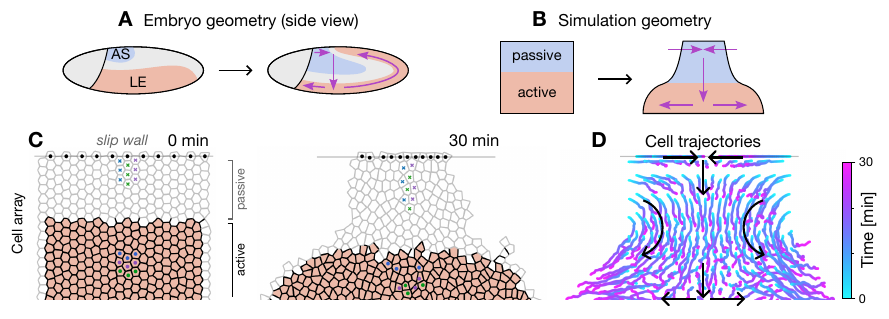}
    \caption{
    \textbf{Combining active and passive tissue regions.}
    \textbf{A, B}~The ellipsoidal geometry of the \textit{Drosophila} embryo (A) is mimicked by a simplified simulation geometry corresponding to an unrolled cylinder (B), whose azimuthal axis corresponds to the dorso-ventral (vertical) axis of the embryo. The different behavior of the dorsal amnioserosa (AS) and the lateral ectoderm (LE) tissue is represented by the passive and active regions in the simulation domain respectively.
    \textbf{C}~Positive feedback in the active region amplifies an initial DV anisotropy of tension and thus drives extension along the AP axis.  Since the embryo's circumference is fixed (implemented via a slip wall at the dorsal boundary), the passive region is stretched along the DV axis. Only half of the simulation domain is shown, corresponding to one lateral side of the left-right symmetric embryo. Three-by-three patches of cells are highlighted to show cell rearrangements (cf.\ Movie~4).
    \textbf{D}~Trajectories of cell centroids showing the tissue scale flow, resembling the characteristic flow of \textit{Drosophila} germ-band extension \cite{Streichan.etal2018}.
    }
    \label{fig:tissue-model-WT}
\end{figure*}

The epithelium of the early \textit{Drosphila} embryo forms a closed, approximately ellipsoidal surface.
Therefore, the deformation of one tissue region has to be compensated by an opposite deformation elsewhere. Specifically, the dorsal amnioserosa is passively stretched along the dorso-ventral (DV) axis and compressed along the anterior-posterior (AP) axis to compensate the convergence extension of the germ band.
To investigate this interplay of active and passive tissue deformations, we mimic the cylindrical geometry of the embryo's trunk (Fig.~\ref{fig:tissue-model-WT}A) by a rectangular tissue patch with ``slip walls'' at the top and bottom boundary (Fig.~\ref{fig:tissue-model-WT}B).
Along the slip walls cell centroids are restricted to move along the wall, thus fixing the DV extent (i.e.\ ``circumference'') of the tissue.
To account for the different mechanical properties of the lateral ectoderm and the dorsal tissue, we divide the tissue into active and passive regions \cite{Streichan.etal2018,Brauns.etal2024elife}.
In the former, cortical tensions are governed by positive feedback in the active region while tension homeostasis is imposed in the latter.
Further, passive cells (subscript $p$) are taken to be soft $\mu_p = 0.2\mu_a, \; \lambda_p = 0.2\lambda_a $ compared to active cells \cite{Rauzi.etal2015}. In addition, we allow interface angles in the passive region to slightly deviate from those imposed by the tension triangulation, reflecting the fact that the overall scale of cortical tensions is lower in the passive tissue \cite{Streichan.etal2018}. 
We initialize the simulation with a slightly perturbed hexagonal packing of cells and the experimentally observed tension anisotropy aligned along the DV axis \cite{Brauns.etal2024elife}.

Starting from this initial condition, the simulation reproduces salient features of the tissue-scale dynamics in the embryo (see Fig.~\ref{fig:tissue-model-WT}C and Movie~4). In the active region (``lateral ectoderm'', LE) active cell rearrangements drive tissue extension along the AP axis and contraction along the DV axis. The passive region (``amnioserosa'', AS) is stretched along the DV axis, accommodating the fixed circumference of the embryo. Notably, this stretching leads to T1s in the passive region as is visible from the highlighted cells in Fig.~\ref{fig:tissue-model-WT}C.
On the tissue level, the coupling of active and passive regions gives rise to the tissue flow pattern characteristic of \textit{Drosophila} germ-band elongation \cite{Streichan.etal2018} as shown in Fig.~\ref{fig:tissue-model-WT}D.

\subsection*{Tissue extension by active T1s requires large-scale mechanical patterning and cell shape elasticity}

\begin{figure*}[tp]
    \centering
    \includegraphics{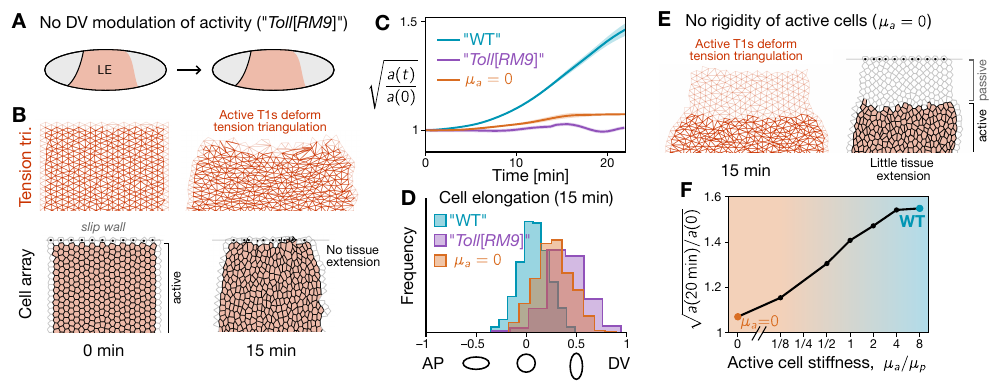}
    \caption{
    \textbf{Tissue flow requires cell rigidity and large-scale genetic patterning.}
    \textbf{A}~Cartoon of an an embryo without dorso-ventral mechanical patterning, such as a \emph{Toll[RM9]} mutant where cells all around the circumference adopt a lateral ectoderm (LE) fate.
    \textbf{B}~Simulation of a tissue patch without a passive region. While active T1s deform the tension triangulation (top), the cell array (bottom) is blocked from elongating by the fixed DV ``circumference'' (implemented via slip walls). Instead, cell rearrangements are compensated by isogonal cell elongation (quantified in D)
    \textbf{C}~CE (measured by the aspect ratio $a(t)$ of the active region) is strongly suppressed in absence of DV modulation of activity (``No DV'') and when active cells have vanishing shear modulus ($\mu_a = 0$; see E and F).
    \textbf{D}~Histograms of cell shape elongation measured by relative difference, $(S_\text{DV,DV} - S_\text{AP,AP})/(S_\text{DV,DV} + S_\text{AP,AP})$, of the AP-AP and DV-DV components of the shape tensor $S$. In the WT case (cf.\ Fig.~\ref{fig:tissue-model-WT}), cells remain nearly isotropic while they become significantly oriented along the DV axis when DV modulation of activity or rigidity of active cells are abolished.
    \textbf{E}~When the active cells have no shear rigidity ($\mu_a = 0$), cell rearrangements are compensated by isogonal cell elongation (see D) without incurring an elastic energy build-up. Thus, almost no tissue-scale CE takes place.
    \textbf{F}~Net amount of CE as a function of the shear modulus of active cells shows that active cells need to be stiffer than the surrounding passive tissue for active T1s to drive efficient CE.
    }
    \label{fig:tissue-model-mutants}
\end{figure*}

The total tissue extension found in the simulations that combine active and passive tissue regions is smaller than the extension of active tissue patches with free boundaries (compare Fig.~\ref{fig:tissue-model-WT}D and Fig.~\ref{fig:tissue-model-disorder}C). 
This suggests that the passive tissue resists deformation.
In the following, we further investigate the role of the spatial modulation of the cells' mechanical properties along the DV axis.
Figure~\ref{fig:tissue-model-mutants}B shows a simulation without DV modulation where all cells are active.
Positive tension feedback drives active T1s everywhere, as is manifest in the deformation of the tension triangulation (Fig.~\ref{fig:tissue-model-mutants}B, right).
However, because of the slip-wall boundary conditions, the tissue cannot contract along the DV axis so that T1s do not result in tissue convergence extension (see Movie~5; quantification in Fig.~\ref{fig:tissue-model-mutants}C).
Instead, cell rearrangements are compensated by isogonal deformations resulting in elongated cell shapes (as quantified in Fig.~\ref{fig:tissue-model-mutants}E).
We predict that this scenario will be realized in \textit{Toll[RM9]} mutant embryos where all cells around the embryo's circumference adopt a ventro-lateral fate \cite{Irvine.Wieschaus1994}, as illustrated in the cartoon in Fig.~\ref{fig:tissue-model-mutants}A.
The stretching of cells leads to a buildup of elastic energy (see Fig.~\ref{SI-fig:cauterized}D).
In \textit{Toll[RM9]} mutant embryos, some of this elastic energy is released by the formation of folds (buckling) \cite{Irvine.Wieschaus1994} which our 2D simulations cannot capture.

Interestingly, the length dynamics of collapsing and emerging interfaces is not significantly affected by the lack of tissue-scale mechanical patterning (Fig.~\ref{fig:tissue-model-mutants}B), even though there is no tissue extension.
This behavior has been observed in experiments and was interpreted as evidence for additional active mechanisms that drive interface elongation, e.g.\ such as medial myosin pulses \cite{Collinet.etal2015}.
However, interface extension in our model is a purely passive process, resulting from the temporal asymmetry of the intercalation process (i.e.\ the low level of active tension on the new interface), we conclude that no such mechanisms are necessary (see Fig.~\ref{SI-fig:cauterized} for further details).

The resistance of cells against shape deformations controls the isogonal modes and therefore is an important parameter controlling the interplay of active and passive tissue deformations. 
Fig.~\ref{fig:tissue-model-mutants}F shows a simulation where the shear modulus $\mu_a$ of active cells is set to zero. Active T1s can therefore be fully compensated by cell elongation through isogonal deformations without incurring an elastic energy cost. As a result, there is no net tissue deformation (see Movie~5).
In other words, cell-shape rigidity is required to maintain rotund cell shapes (i.e.\ resist isogonal shear deformations) and thus translate active T1s into net tissue deformation.
The tissue deformation by isogonal modes is determined by a balance of external forces and internal resistance of cells to shape changes.
Here, the external forces acting on the active tissue result from the passive tissue's resistance to deformation, which is in turn set by shear modulus $\mu_\mathrm{p}$.
The ratio of the shear moduli in the active vs the passive region, $\mu_\mathrm{a}/\mu_\mathrm{p}$, determines how much the active region deforms (see Fig.~\ref{fig:tissue-model-mutants}G).
Only when the cells in the active tissue are more rigid than those in the passive region ($\mu_\mathrm{a}/\mu_\mathrm{p} > 1$), is it energetically favorable to isogonally deform the passive region rather than the active region. 
This predicts that GBE can be impaired by stiffening the dorsal tissue (amnioserosa), e.g.\ in \textit{dpp} mutants.

\section*{Discussion}

Our model provides a theory of active elasticity based on the geometric relation (duality) between tension space and real space afforded by the force balance condition.
Stabilizing feedback mechanisms that maintain adiabatic force balance are implicit in our model as we constrain the tension dynamics to the space of force-balanced configurations (flat tension triangulations).
Dynamics in tension space is driven by local positive feedback.
This feedback amplifies a weak initial tension anisotropy and thus drives cell shape dynamics that result in cell rearrangements (T1 processes).
Notably, while active T1s are initiated by positive tension feedback, we find that their resolution is through passive relaxation of tension and does not require additional active ingredients as previously suggested \cite{Collinet.etal2015}.
Force balance provides the non-local coupling that allows for the coordination of forces and cellular behaviors across the tissue.
On the tissue scale, self-organized active T1s are oriented by global tension anisotropy and thus act coherently to drive convergent-extension flow.
As T1s drastically remodel tension geometry, they gradually degrade the orientational cue provided by initial tension anisotropy.
(Note that this degradation of order happens locally through T1 transitions; its relation to long-wavelength instabilities of active nematics in which there are no explicit topological transitions is an interesting avenue for future research.)
Thus, the locally self-organized tissue flow is generically self-limiting. 
It arrests after a finite extent of convergent extension (CE) that depends on the initial degree of order in the cellular packing and the magnitude of initial tension anisotropy.
This central finding suggests that cell geometry is a repository of morphogenetic information that may encode the final tissue shape.

Importantly, we show that T1s are controlled by local configuration of tensions as quantified by the LTC order parameter that links the locally hexatic space of the tension triangulation (representing force balance) with the nematic nature of deviatoric stress and strain in physical space. 
In contrast to previous work on ``hexanematic'' order in tissues \cite{Armengol-Collado.etal2023,Armengol-Collado.etal2024}, LTC order is defined in tension space, not in physical space, and does not require (quasi-) long-range hexatic order. Moreover, the emergence of this order is not driven by a free energy but by active feedback acting on tensions as discussed further below.

Mechanically self-organized tissue dynamics provides an elegant explanation for the arrest of \textit{Drosophila} GBE after about two-fold elongation \cite{Irvine.Wieschaus1994}, while predicting its dependence on the initial tension anisotropy \cite{Streichan.etal2018,Gustafson.etal2022}.
In the companion paper \cite{Brauns.etal2024elife}, we used tension inference and LTC analysis to re-examine live imaging data on \textit{Drosophila} germ-band extension (GBE) \cite{Stern.etal2022} showing that it is driven predominantly by the internally generated forces in the lateral ectoderm of the embryo.  Quantitative analysis of imaging data confirms that the \textit{Drosophila} ectoderm starts ordered and becomes disordered as T1s proceed during CE.

Notably, while our model does not require cell-scale genetic instructions to generate local tension anisotropy\cite{Irvine.Wieschaus1994,Pare.etal2014}, genetic patterning on the scale of the embryo is essential for the coordination and stability of the global flow. 
In the early \textit{Drosophila} embryo, this is manifested in the dorso-ventral patterning system that specifies the tissues with different mechanical properties and modulates mechanical feedback loops \cite{Gustafson.etal2022}.
We expect that the initial tension anisotropy, reported in \cite{Brauns.etal2024elife}, is set up by anisotropic static ``hoop'' tension resulting from turgor pressure inside the embryo, and further reinforced by the dynamic effects of ventral furrow formation \cite{Gustafson.etal2022}.
An important challenge for future work is to identify the (molecular) mechanisms of both positive and negative feedback circuits, the latter stabilizing the force-balanced configuration on short timescales while the former, driving controlled remodeling on long timescales.

Our model predicts that disrupting the hexagonal packing of nuclei prior to cellularization will cause slower GBE.
Interesting candidates to test this prediction are ``nuclear fallout'' mutants, where some nuclei leave the blastoderm surface and thus introduce defects in the cellular packing \, cite{Rothwell.etal1998}. Another option might be the transient and partial disruption of microtubule organization with small molecule inhibitors \cite{Kanesaki.etal2011}.
We expect that these experiments can be used to challenge and subsequently refine the model.

Comparing the LTC time courses between experiments and simulations, we find an excellent agreement, suggesting that positive feedback-driven local tension dynamics can explain cell-scale behavior during GBE. 
The dynamics in tension configuration space depends on the character of the positive tension feedback. Winner-takes-all feedback efficiently drives the local tension configuration toward the T1 threshold via the formation of tension bridges. By contrast, when feedback saturates at too low relative tension, it causes the formation of tension cables, which had previously been suggested as a driver for CE.
However, our simulations and analysis of local tension configurations show that tension cables are inefficient at driving CE, as adjacent interfaces ``compete'' to contract.
Indeed, the arrest of CE due to the formation of tension cables is also observed in recent computational studies \cite{Sknepnek.etal2023,Rozman.etal2023}. 
When tension cables contract, they lead to the formation of ``rosettes'' where five or more cells meet in a single vertex \cite{Zallen.Zallen2004,Blankenship.etal2006}.
In a \textit{Drosophila} mutant for \textit{abl}, contraction of tension cables is impaired resulting in a reduction of rosette formation \cite{Tamada.etal2012}. By contrast, T1s appear unaffected in these mutants, suggesting that isolated high-tension junctions contract normally.
Our model offers a possible explanation of this puzzling finding: Deletion of abl might cause the positive tension feedback to saturate earlier thus leading to the formation of persistent tension cables. 
Taken together, the findings discussed above show that tension bridges are key to driving efficient convergent-extension flow.
The LTC order parameter introduced here facilitates statistical analysis across many cells and allows one to distinguish different regimes of local tension dynamics. 

Because force balance geometry does not uniquely define the shape of cells -- on account of isogonal degrees of freedom -- the latter play an important role in defining tissue dynamics.  
Isogonal soft modes account for tissue deformation under constant cortical tensions and are controlled by non-cortical mechanical stresses, arising e.g.\ from passive cell elasticity due to cell-internal structures (e.g.\ nucleus \cite{Grosser.etal2021,Kim.etal2022nuclear}, microtubules, and intermediate filaments \cite{Pensalfini.etal2023}).
We find that internal rigidity is essential to transduce cell intercalations into tissue-scale deformation against resistance from adjacent tissues. In the absence of cell resistance against deformation, intercalations are compensated by cell shape changes. 
However, if cell-internal elasticity becomes stronger than cortical tensions, it can resist changes in vertex angles and impede T1s and tissue flow
(Fig.~\ref{SI-fig:epsilon_dependence}). Such a scenario may occur in certain genetic mutants, like the ``kugelkern'' (\textit{kuk}) mutant of \textit{Drosophila}, where the nucleus is stiffer \cite{Brandt.etal2006}).

Overall, our findings suggest that epithelial tissue flows not like a fluid (where the shear modulus vanishes) but rather as a plastically deforming solid, whose remodeling is driven internally, while resisting external forces. Epithelial tissue can thus be regarded as an \emph{active solid}.
More generally, tissue mechanics dominated by cortical tensions that are controlled by feedback mechanisms call for a formulation of continuum mechanics that does not rely on a constitutive stress-strain relationship.

\acknowledgements{We thank Arthur Hernandez, Matthew Lefebvre, Noah Mitchell, Sebastian Streichan, and Eric Wieschaus for stimulating discussions. We further thank Dinah Loerke and Jennifer Zallen for their insightful feedback.
FB acknowledges support of the GBMF post-doctoral fellowship (under grant \#2919). NHC was supported by NIGMS R35-GM138203 and NSF PHY:1707973. BIS acknowledges support via NSF PHY:1707973 and NSF PHY:2210612.}

\bibliography{GBE}

\clearpage

\onecolumngrid

\newgeometry{total={6in,8in}}

{\centering\LARGE\scshape Supplementary Material \par}

\setcounter{section}{0}
\renewcommand{\thesection}{\arabic{section}}
\renewcommand{\thesubsection}{\arabic{section}.\arabic{subsection}}

\setcounter{figure}{0}
\renewcommand{\thefigure}{S\arabic{figure}}

\setcounter{equation}{0}
\renewcommand{\theequation}{S\arabic{equation}}

\section{Analysis of experimental data}

\subsection{Experimental data source}

Whole embryo cell segmentation and tracking data was obtained from the repository deposited with Ref.~\cite{Stern.etal2022}: \href{https://figshare.com/articles/dataset/Deconstructing_Gastrulation_-_Data/18551420/3}{DOI:10.6084/m9.figshare.18551420.v3}.
We analyzed the dataset with ID number 1620 since it had the highest time resolution (\SI{15}{s}) and covered the longest time period (\SI{50}{\minute}), starting ca.\ \SI{7}{\minute} before the onset of ventral furrow invagination.

\subsection{Hexagonal packing and hexatic order}

\begin{figure}[t]
    \centering
    \includegraphics{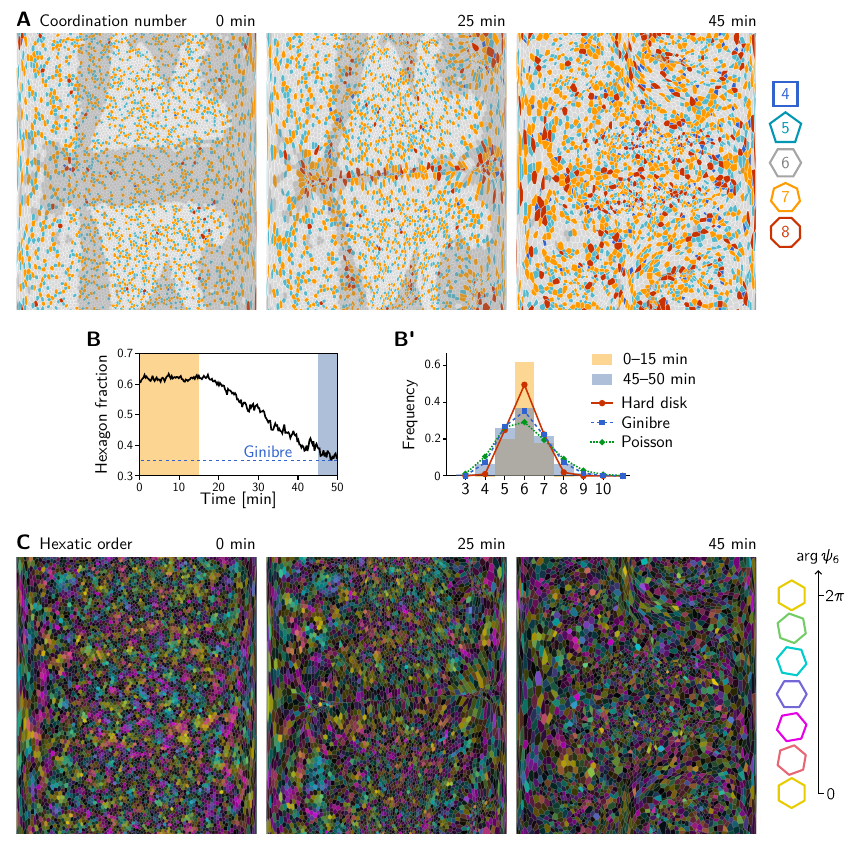}
    \caption{Coordination number and hexatic order of cells.
    \textbf{A}~Initially (0~min), the majority of cells are hexagonal (light gray). During GBE (25--50~min), topological defects proliferate, resulting in increasing numbers of non-hexagonal cells. Semitransparent gray overlay marks cells that invaginate. 
    \textbf{B} and \textbf{B'}~Fraction of hexagonal cells as a function of time (B) and histogram of cell-coordination number (B') in the lateral ectoderm. The histograms show pooled data from the time periods 0--15~min and 45--50~min highlighted in B. Lines indicate the coordination number distributions for Voronoi tessellations generated from three different random point processes: hard disk (packing fraction 0.54 near the theoretical maximum for sequential placement \cite{Feder1980}), Ginibre \cite{Ginibre1965}, and Poisson. 
    \textbf{C}~Hexatic order is low, and shows no long-range correlation. It further decreases as GBE progresses. Hue and brightness code for the phase ($\mathrm{arg} \, \psi_6$) and magnitude ($|\psi_6|$) of the hexatic order parameter, respectively. 
    }
    \label{SI-fig:hexagonal-order}
\end{figure}

We quantify the order of the cell packing in the tissue by two measures: (\textit{i}) the fraction of six-sided cells which one might loosely call ``hexagonal packing order'' and (\textit{ii}) the hexatic order parameter measuring bond-orientational order.
Hexagonal packing order is a topological quantity: it depends only on the neighborhood relations between cells but not on their exact shape. In contrast, hexatic order (defined below) is a geometric measure sensitive to the cell shapes. 
Initially, the majority of cells have six neighbors (Fig.~\ref{SI-fig:hexagonal-order}A, 0~min), and the density of topological defects, manifested as non-hexagonal cells, is relatively low. 
During VF invagination, the number of defects increases only slightly (25~min) since there are only very few intercalations \cite{Brauns.etal2024elife}.
During GBE, the number of defects increases significantly due to T1s. 
Towards the end of GBE, the distribution of cell-coordination number approaches that of a random Voronoi tessellation seeded with a Ginibre random point process (see Fig.~\ref{SI-fig:hexagonal-order}B and B').

\paragraph*{Hexatic order.} The hexatic order parameter (also called ``bond orientational order parameter'' \cite{Halperin.Nelson1978}) for a cell with $n$ vertices is defined as
\begin{equation}
    \psi_6 = \frac{1}{n} \sum_{k = 1}^{n} \exp (6 i \theta_k),
\end{equation}
where $\theta_k$ is the angle between the DV axis and the vector pointing from the cell's centroid to vertex $k$.
For a cell with a regular hexagonal shape, the magnitude of $\psi_6$ reaches its maximum $|\psi_6| = 1$. The phase $\mathrm{arg} \, \psi_6$ indicates the orientation of the hexagon.
In contrast to the coordination number, which is a purely topological measure, the hexatic order parameter depends on the cell shape.
In line with previous findings from confocal microscopy of the \emph{Drosophila} germ band \cite{Zallen.Zallen2004}, we find that the hexatic order parameter is low in magnitude and exhibits no long-range correlations as is apparent from Fig.~\ref{SI-fig:hexagonal-order}C.
Effectively, the presence of non-hexagonal cells acts as an obstruction to long-range correlations of geometric order. 
We quantify the range of correlations in hexatic order by coarse-graining over patches of cells with different radii (measured by the neighborhood degree).
Before the onset of GBE, the coarse-grained order parameter decays as a power law of the patch radius law with an exponent that is close to $-0.75$, a value found also for MDCK cells and in simulations using multiphase-field models \cite{Armengol-Collado.etal2023}.
(We find the same result using the distance-weighted hexatic order parameter introduced in Ref.~\cite{Armengol-Collado.etal2023}.)
At the end of GBE, the decay exponent is smaller than $-1$, indicating a complete lack of correlations in hexatic order between neighboring cells. This is expected from the high number of non-hexagonal cells at this stage (Fig.~\ref{SI-fig:hexagonal-order}B, 50~min).

\subsection{Local and regional tension inference}

\begin{figure}[t]
    \centering
    \includegraphics[width=\textwidth]{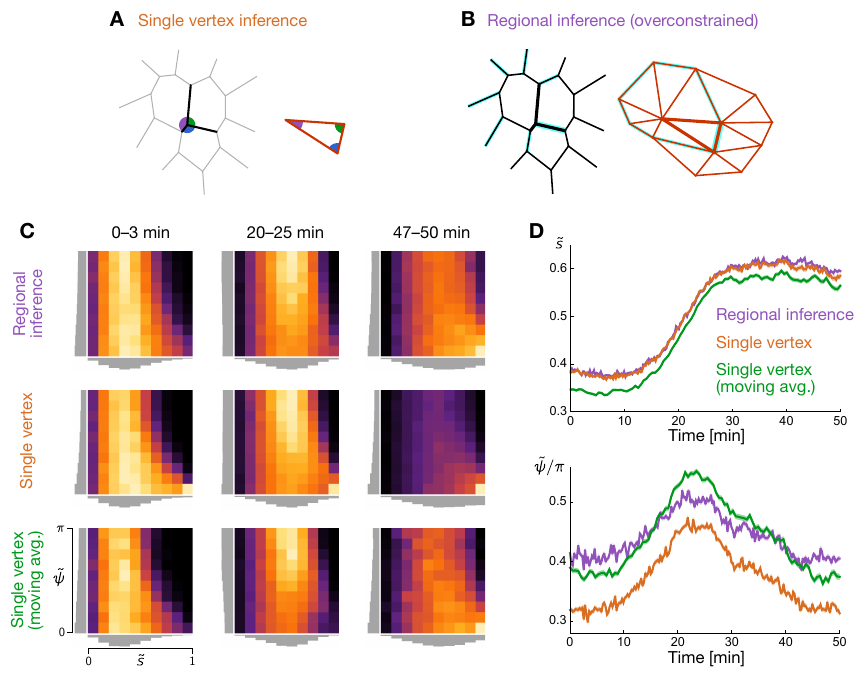}
    \caption{
    Comparison of local vs regional tension inference.
    \textbf{A}~Local tension inference at a single vertex through the complementarity of angles at a single vertex (left) and in the corresponding tension triangle (right). The relative tensions (edge lengths of the tension triangle) are found via the law of sines. 
    \textbf{B}~Regional tension inference for a patch of three cells. The interfaces (and corresponding tensions) shared between them are highlighted by bold lines. The triangulation property of the tensions yields one additional constraint for each cell because for each cell there is a loop in the tension triangulation which has to close. One such loop is highlighted. These additional constraints make the regional inference more robust.
    \textbf{C}~Comparison of the LTC distributions for regional and local tension inference in the \textit{Drosophila} germ band. Distributions are from data collected over the indicated time intervals. Germ band elongation starts at ca. 25~min. The last row shows single-vertex inference based on vertex positions smoothed using a moving average (1.25~min interval).
    \textbf{D}~Mean anisotropy magnitude (top) and LTC phase (bottom) for the three inference variants. Note the systematically lower LTC phase, i.e.\ lower fraction of tension bridges, found using single-vertex inference.
    }
    \label{SI-fig:local-vs-regional-inference}
\end{figure}

\begin{figure*}[tp]
    \centering
    \includegraphics{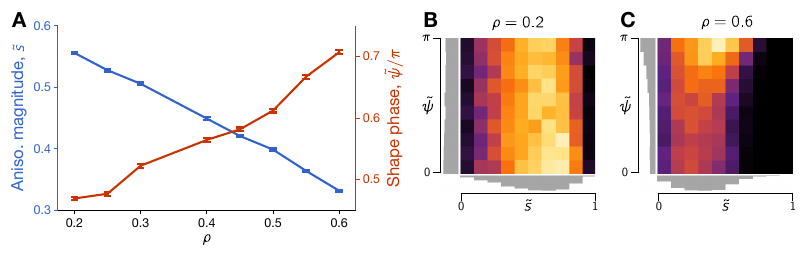}
    \caption{LTC distribution for the hard-disk Delaunay family of random triangulation.
    \textbf{A}~Median LTC parameter magnitude and phase vs hard disk packing fraction, $\rho$. 
    \textbf{B-C}~LTC histograms for two different packing fractions show significant differences in the LTC distribution. Note that topological statistics such as coordination numbers are much less sensitive to $\rho$.
    LTC distributions were computed from cell arrays with $4000$ cells.
    }
    \label{SI-fig:LTC-Random-Delaunay}
\end{figure*}

To quantify the local tension configurations in experimental data from the \textit{Drosophila} germ band, we performed tension inference based on segmented cell outlines. 
In the companion paper \cite{Brauns.etal2024elife}, tension inference was done locally, directly relating the angles at each vertex to the relative cortical tension via the law of sines (see Fig.~\ref{SI-fig:local-vs-regional-inference}A). 
While this method is conceptually and computationally simple, it is sensitive to observational and dynamical noise in the angles.
A more robust approach is to perform tension inference in an extended region, which makes the inference problem overconstrained \cite{Noll.etal2017}. In particular, there is one additional constraint per cell because the tension triangles for each cell have to fit together in force balance as illustrated in Fig.~\ref{SI-fig:local-vs-regional-inference}B.
The overconstrained inference returns tensions for the closest force-balanced configuration compatible with the observed angles and thus removes small deviations from tensional force balance due to pressure differentials, short-time fluctuations, and observational noise.
(An alternative method to remove observational noise is to apply a moving average on the vertex positions before inferring tensions.)
Since in our model cortical tensions are always in exact tensional force balance, we use this overconstrained inference to compare to simulations. Specifically, we use tension inference on all interfaces of the three cells that meet at a given vertex to determine the local tension configuration parameter at that vertex.
Figure~\ref{SI-fig:local-vs-regional-inference}C shows the LTC distributions obtained from local and regional inference. The LTC distributions are qualitatively very similar but they differ quantitatively. In particular, the LTC phase $\tilde{\psi}$ has the same qualitative trend with a transient increase before the onset of T1s but is systematically lower for regional inference.

\section{Additional simulation results} 
\label{SI:additional-simulations}

In this section, we report additional results from the tissue scale simulations and/or explain implementation details.

\subsection{Dynamics of the shape index}

A common measure for the degree of order in a polygonal tiling is the cell shape index ($s = P/\sqrt{A}$, with cell perimeter $P$ and area $A$), shown in Fig.~\ref{SI-fig:disorder}A.
This shape index is high when cell shapes are elongated or irregular and approaches the minimal value $s \approx 3.72$ for a regular hexagon.
The shape index has particular relevance in vertex models employing an area--perimeter elastic energy \cite{Hufnagel.etal2007,Farhadifar.etal2007}.
In these models, the \emph{target} shape index is a control parameter that drives a solid-to-fluid transition \cite{Bi.etal2015}: 
a high target shape index ($s_0 > s_0^* \approx 3.81$) is associated with tissue fluidity since it allows for cell rearrangements, while a low cell target shape index gives rise to a solid state. In both cases, the observed shape index is controlled by the target shape index.
In contrast, in our simulations with actively driven T1s, disorder, and thus a high observed shape index, is the consequence of cell rearrangements, rather than their cause.
Notably, we find more tissue flow when the observed shape index is initially low (Fig.~\ref{SI-fig:disorder}A).
The question of the solid vs fluid character of tissue is addressed in more detail in the last part of the results section and in the discussion. 

\subsection{Dynamics in order-anisotropy-extension phase space}

As we have seen in the main text, both order and anisotropy decrease as cells rearrange (cf.\ Fig.~\ref{fig:tissue-model-disorder}E).
When plotting the remaining extent of convergent extension $\sqrt{a_\mathrm{final}/a(t)}$ against $p_6(t)$ and $||\langle \sqrt{2}\tilde{Q}(t) \rangle||$, we find trajectories that approximately lie in a common plane and converge to a fixed point at vanishing anisotropy $||\langle \sqrt{2}\tilde{Q} \rangle|| \approx 0$ and $p_6 \approx 0.4$ (Fig.~\ref{SI-fig:disorder}B). 
Based on these results, we hypothesize an empirical law for feedback-driven convergent extension, based on the instantaneous hexagon fraction and tension anisotropy:
\begin{align}
    \label{eq:linear_fit_extension}
   \sqrt{\frac{a_\mathrm{final}}{a(t)}} \approx c_Q ||\langle \sqrt{2} \tilde{Q}(t) \rangle|| + c_6 p_6(t)
\end{align}
With the coefficients determined by a linear fit, \eqref{eq:linear_fit_extension} predicts the remaining extent of convergent extension with less than 10\% mean absolute error (Fig.~\ref{SI-fig:disorder}C).
The empirical relation \eqref{eq:linear_fit_extension} is specific to the particular choice of microscopic tension dynamics employed in the simulations:
a systematic search for a more general relation linking microscopic ($p_6, ||\langle \sqrt{2} \tilde{Q}(t) \rangle||$) and macroscopic ($a(t)$) quantities is beyond the scope of the present work.

\subsection{Influence of isogonal stretching on the LTC distribution}

Here, we consider the influence of the reference cell shape $S_0$ which determines the isogonal potential via the elastic energy. As discussed in the main text, isogonal stretching or compression of cells along the axis of tension anisotropy can delay or accelerate T1s by moving the T1-threshold in LTC space. As reported in the main text, if we implement isogonal stretching by choosing an anisotropic $S_0$, i.e.\ $S_0 = 3 \ell_0 \; \mathrm{diag}(1-t, 1+t), \; t\in [0, 1]$, we find that a higher bridge bias emerges at the early phase of convergent extension (Fig.~\ref{fig:tension-config-data}). We chose an anisotropic reference shape  $t=1/3$ to model the isogonal stretching caused by the ventral furrow before the onset of GBE. In the experimental data, this isogonal stretch decays; analogously, we linearly ramp the reference shape anisotropy down so that $t=0$ at 50 minutes simulation time.
We note that our modeling of isogonal stretching is limited since it is encoded in a model parameter ($S_0$) instead of being created dynamically, for example by external forces applied to the boundary.
Time traces of the LTC parameter in the absence of isogonal stretching are shown in Fig.~\ref{SI-fig:parameter_scan}) below.

\subsection{Alternative forms of positive tension feedback}

We now turn to discussing two variants of the triangle-intrinsic tension dynamics. In the main text, we discussed two types of positive feedback, saturating and winner-takes-all. In both cases, the overall scale of tensions was determined by keeping the triangle perimeter $P=\sum_{\alpha\in I_{ijk}} \tilde{T}_\alpha$ constant. This corresponds to a fixed amount of total active tension that is only redistributed across edges. We can also consider a model where the triangle area remains fixed. This can be implemented using the gradient $(J_A)_{ij} = \partial_{\tilde{T}_{ij}} A_{I_{ijk}}$ of the triangle area with respect to the triangle side lengths:
\begin{align}
    \partial_t \tilde{T}_{ij} &= {\tau_\mathrm{T}}^{-1} \left(\tilde{T}_{ij}^{n} - \frac{1}{\sqrt{\sum_{\alpha \in I_{ijk}} ( \partial_{\tilde{T}_{\alpha}} A_{I_{ijk} } )^2} } \sum_{\alpha\in I_{ijk}} (\partial_{\tilde{T}_{\alpha}} A_{I_{ijk}} \tilde{T}_\alpha^n) \right)
    \label{eq:tension-dynamics-area}
\end{align}
The overall dynamics of this model are very similar to \eqref{eq:tension-dynamics} considered in the main text. However, the tension feedback is ``more aggressive'' since now the total tension (triangle perimeter) increases as the tension triangle becomes more anisotropic. This mirrors a situation where in addition to the anisotropy also the total myosin levels increase during GBE \cite{Lefebvre.etal2022}. Therefore, one observes slightly larger amounts of convergent extension for identical initial conditions.

Next, we considered adding small amounts of i.i.d.\ Gaussian noise to the cortical tension dynamics, i.e.\ a stochastic, Langevin tension evolution
\begin{align} \label{eq:tension-dynamics-stochastic}
    &\tau_\mathrm{T}^{} \partial_t \tilde{T}_{ij} = \tilde{T}_{ij}^{n} - \frac{1}{3} \sum_{\alpha \in I_{ijk}} \tilde{T}_{\alpha}^n +\eta_{ij}, \\
    &\text{with} \; \eta_{ij} \sim_\text{i.i.d.} \text{Normal}(0, \sigma^2) \; \text{and} \; \langle \eta_{ij}(t) \eta_{kl}(t') \rangle = \sigma^2 \delta(t-t') \delta_{ij, kl}
\end{align}
To integrate \eqref{eq:tension-dynamics-stochastic}, we use the explicit Euler-Maruyama scheme. We find that the convergent extension phenomenology is robust to low to moderate levels of noise (i.e.\ $\sigma < T_{ij}^n$). However, higher $\sigma$ leads to a final LTC distribution with somewhat more anisotropic triangles, indicating larger disorder, and reduces the amount of total convergent extension.

Cell arrays from a Langevin simulation and from a simulation with an area-based myosin pool mechanism are shown in Fig.~\ref{SI-fig:alternative_models_meshes}. A quantitative analysis is shown in Fig.~\ref{SI-fig:parameter_scan}.

\subsection{Saturating tension feedback}

Here, we present the details for the simulations of saturating tension feedback in main text Fig.~\ref{fig:saturating-feedback}. We consider bistable tension dynamics of the form
\begin{align}
    \tau_T \frac{d}{dt} \dot{T} = -(T-T_-)(T-T_c)(T-T_+)
\end{align}
where $T_-<T_c<T_+$ are the low, unstable, and high tension fixed points, respectively. We set $T_-=0, \; T_c=1$ matching the fixed points of the main feedback model \eqref{eq:tension-dynamics} we consider in this manuscript. The simulations with saturating feedback correspond to $T_+=1.2$, the control simulations are $T_+=3.33$. Note that we adjusted the time step and all other rate parameters of the simulation so that $\dot{T}$ is similar across different choices of $T_+$.

\paragraph*{Irreversible and reversible T1 transitions.}

For saturating tension feedback, we observe that approximately 20\% of T1s reverse, i.e.\ the newly formed junction re-collapses. For winner-takes-all feedback, less than 1\% of T1s reverse.
Reversing T1s are defined as follows. Consider an edge $ij$ between cells $i$ and $j$, the central edge in a quartet of cells $ikjl$ (in clockwise order).
The collapse of edge $ij$ creates an edge $kl$. This T1 is considered reversible if $kl$ collapses, in turn, to give rise to a connection $ij$ again, i.e.\ $kl$ re-collapses before any of the outer edges $il$, $lj$, $ik$, or $ki$, of the quartet have collapsed. In this case, the local quartet topology returns to its original state. 
We also filter out nested sequences of multiple reversible T1s, but such events are extremely rare.

\subsection{Effect of hexatic order orientation}

When carrying out simulations in which the tissue is initially ordered, the initial tension triangulation corresponds to a (rectangular) patch of a triangular lattice. We can choose two possible orientations of the lattice plane with respect to the tension anisotropy -- one creating initially mainly tension cables, one mainly bridges. This corresponds to the orientation of the hexatic order $\Psi_6$ w.r.t. the axis of tension anisotropy, and we refer to these orientations as ``bridge'' and ``cable'' initial conditions. We find that the qualitative behavior of the model is independent of this choice, and positive tension feedback is capable of creating coherently ordered T1s even if future collapsing edges are not singled out in the initial condition.  To underline this point, Fig.~\ref{fig:tissue-model-disorder} uses ``cable'' initial conditions. For the determination of LTC distribution dynamics in main text Fig.~\ref{fig:tension-config-data}, as well as the heat map Fig.~\ref{fig:tissue-model-disorder}D, we pool simulations from cable and bridge initial conditions. When showing snapshots from a simulation, the  ``parallel'' initial condition most easily creates the recognizable pattern of alternating high and low tensions reported in the companion paper.

A quantitative comparison is shown in Fig.~\ref{SI-fig:cable_vs_bridge_initial}. We see that an initial bridge bias leads to a higher bridge fraction later on, and somewhat larger amounts of convergent extension.

\subsection{Interface extension is independent of tissue extension}

As argued in the main text, interface extension after active T1s is not driven by external forces (e.g.\ due to tissue elongation) but locally by cortical tensions. Active T1s are fundamentally asymmetric in time: the collapsing interface has high myosin levels, and the newly formed interface has very low levels. Since these low levels are insufficient to balance the tensions of the surrounding interfaces, the new interface extends. 

Our simulations support the notion that interface extension does not depend on tissue extension, even though our simulations contain no additional active mechanisms to elongate new interfaces. In the main text, we showed simulations of a tissue geometry without a soft, passive region to compensate for the convergence of the active tissue (analogous to a \emph{Toll[RM9]} mutant). Because of incompressibility, the active tissue also does not extend. Nevertheless, new interfaces elongate at rates very similar to control simulations (Fig.~\ref{SI-fig:cauterized}B). We further support this conclusion with an additional simulation. Mimicking experiments where tissue extension is blocked by a cauterization fence \cite{Collinet.etal2015}, we block tissue extension by adding slip walls along the AP boundaries of the tissue patch (Fig.~\ref{SI-fig:cauterized}A). For numerical reasons, we add a small passive zone in front of these additional boundaries and enforce the slip-wall condition with a softer penalty than for the dorsal slip walls. This is done to avoid direct conflicts between the boundary conditions and the angle constraints from the triangulation. In these simulations, the active tissue does not extend, and cells deform to compensate T1s (Fig.~\ref{SI-fig:cauterized}A). However, new interfaces still successfully extend (Fig.~\ref{SI-fig:cauterized}B), like in the \emph{Toll[RM9]} simulations shown in Fig.~\ref{fig:tissue-model-mutants}. The overall speed of intercalations is slightly reduced because of the additional slip walls.
Note that occasionally, we observe rupture of the simulated tissue from the simulation boundary because of the internal forces generated by T1 transitions. Similar rupture events are seen in experiments.

\subsection{Scale separation between cortical tensions and cell shape elastic energy}

As discussed in the main text, we have formulated a model of epithelial tissue mechanics in the limit where the cortical tensions are the dominant source of stress. This separation of scales is encoded by the small parameter $\varepsilon$ in \eqref{eq:total-energy}.
To lowest order in $\varepsilon$, force balance of the cortical tensions imposes geometric constraints on angles under which the subdominant ``cell-shape energy'' is relaxed.
In other words, the tension-driven and isogonal degrees of freedom are clearly separated.
To higher order in $\varepsilon$, the cell-shape energy changes the angles, thus mixing the tension-driven and isogonal modes.
How small does $\varepsilon$ have to be such that we can ignore this higher-order effect?

To answer this question, we need to compare the tension force balance to the force on vertex angles exerted by the cell-shape energy $\varepsilon \sum_i E_\mathcal{C}(\{\mathbf{r}_{ijk}\})$ (defined in \eqref{eq:cell_shape_energy}).
This force is given by the elastic energy gradient $\nabla E_\mathcal{C} = \partial_{\mathbf{r}_{ijk}} \sum_i E_\mathcal{C}(\{\mathbf{r}_{ijk}\}) $ at the vertex positions $\mathbf{r}_{ijk}^* = \mathrm{argmin}_{\mathbf{r}_{ij} \perp \mathbf{t}_{ij} } \sum_i E_\mathcal{C}(\{\mathbf{r}_{ijk}\})$ which minimize the elastic energy under the angle constraints imposed by the tension triangulation (i.e.\ the tension triangulation edges $\mathbf{t}_{ij}$ and the cell edges $\mathbf{r}_{ij}$ are orthogonal).
At this minimum, a displacement $\delta \mathbf{r}_{ijk} = -h \nabla E_\mathcal{C}(\mathbf{r}^*)$ for a small $h>0$ has no isogonal component and acts purely on angles. We impose this displacement and calculate the tension- and cell-elasticity energy cost via the non-dimensional ratio
\begin{equation} \label{eq:angle_force_ratio}
    f = \frac{\sum_i E_\mathcal{C}(\{\mathbf{r}_{ijk}^* + \delta\mathbf{r}_{ijk} \})-  E_\mathcal{C}(\{\mathbf{r}_{ijk}^*\}) }{\sum_{ij} T_{ij} [\ell_{ij}(\{\mathbf{r}_{ijk}^* + \delta\mathbf{r}_{ijk} \}) - \ell_{ij}(\{\mathbf{r}_{ijk}^*\})]}.
\end{equation}
This expression on data from the simulation shown in Fig.~\ref{fig:tissue-model-disorder}A, we find $f \approx -3 \mu \overline{\ell}/\overline{T}$, where $\overline{\ell}$ and $\overline{T}$ are the average interface length and tension, respectively.
They set the basic units of length and force. In our simulations, we set $\overline{T}=1$ and $\overline{\ell}=1/\sqrt{3}$ (this ensures that the cell sizes are compatible with the Voronoi dual of the tension triangulation \cite{Brauns.etal2024elife}).
$f$ is negative because the cell-shape energy can relax further by changing the angles, compared to when it is constrained to the isogonal (fixed-angle) subspace.
Notably, only the shear modulus $\mu$ of the cells contributes, indicating that the effect of bulk elasticity on the angles is negligible.
The numerical factor of $3$ is a consequence of the particular way we have defined the elastic energy and is not fundamental.
Given our choice of units, the parameter setting $\mu=1$, we must have $\varepsilon \lesssim 0.2$ for the separation of tension-driven and isogonal modes to be valid.

To test this, we numerically ran simulations with different values of $\varepsilon$, $0.001$ (our default value), $0.01$, $0.1$, and $1$. As shown in Fig.~\ref{SI-fig:epsilon_dependence}, we find almost identical behavior of the simulations for $\varepsilon=0.001$ and $\varepsilon=0.1$.
By contrast, for $\varepsilon=1$ tissue shape change is impaired since the tensions are insufficient to overcome the tendency of the cell-shape energy to drive cells to regular hexagon shapes.
In other words, the cell-shape energy imposes energy barriers that prevent T1 transitions. In particular, we find that at $\varepsilon=1$, more than half of T1 transitions reverse, and are hence ineffective, while this is the case for less than $2\%$ of T1s at $\varepsilon=0.001$. Without explicit saturation of the positive tension feedback, this suppression of T1s leads to a buildup of extreme tension anisotropy, eventually leading to numerical problems.

\section{T1-threshold in LTC space}

To find the resulting T1 threshold in the LTC space, we need to express the tension $T$ on the cell quartet's inner interface in terms of the angles $\alpha, \alpha', \beta, \beta'$.
Reasoning that on average the two tension triangles in a quartet will be similar, we consider the simplified case $\alpha = \alpha', \beta = \beta'$.
Normalizing by the average tension we find $T = 3 \sin(\beta)/ (\sin \alpha + \sin \beta+ \sin (\alpha + \beta))$.
Since the problem is now reduced to the shape of one triangle, the T1 condition $\ell_\mathrm{ref}(\alpha, \beta) + \Delta \ell_\mathrm{iso} = 0$ defines a threshold line in the LTC space.
Figure~\ref{fig:tension-config-space}E shows the T1 threshold as a function of the isogonal strain $\Delta \ell_\mathrm{iso}/\ell_0$.
The critical tension anisotropy drops to zero as $\Delta \ell_\mathrm{iso}/\ell_0$ approaches $-1$, the critical isogonal strain for purely passive T1s which takes place for isotropic tension (i.e.\ equilateral tension triangles).
Vice versa, positive isogonal strain shifts the T1 threshold to a higher magnitude of tension anisotropy.
In principle, the above geometric reasoning can be generalized to an arbitrary tension ``kite'' composed of two different tension triangles. However, the shape space of such kites is four-dimensional (since there are four independent angles $\alpha, \alpha', \beta, \beta'$) which precludes the intuitive visualization that the single-triangle LTC space provides.

\section{Isogonal shear and tissue shear modulus}
\label{SI:iso-shear}

As discussed in the main text, and shown in Fig.~\ref{fig:tissue-model-mutants} our simulations show that a non-zero cell-level shear modulus is required for extension via active T1s when the tissue extends against external restoring force (e.g. from adjacent tissues). Here, we analyze the relation between cell-level and tissue-level shear modulus. 
Crucially, because of the dominance of cortical tensions, a tissue patch in our model will respond to externally applied forces via an isogonal deformation. We first show that an isogonal deformation can create tissue-scale shear (on the level of cell centroid displacement -- the transformation of cell vertices is necessarily non-affine to preserve angles). Then we show that these shear modes correspond to the response of the tissue to external force by computing the Hessian of our cell elastic energy in the subspace spanned by isogonal deformations, and measure the shear modulus. Our discussion here applies to deformations of the \emph{bulk} of a tissue; special care is needed when considering deformations that change the tissue boundary, as we explain below.

\subsection{Quadratic isogonal mode profile creates shear deformation}

To show that the isogonal modes can cause shear deformations, not just dilation/contraction of cells, we make use of the isogonal mode parametrization introduced in Ref.~\cite{Noll.etal2017}. It assigns an isogonal ``potential'' $\Theta_i$ to each cell, and calculates the cell displacements from the $\Theta_i$ and the edge tension vectors.
In the following, we will show that a constant gradient in the isogonal potential generates a uniform translation in real space. By integration, this implies that a quadratic spatial profile of the isogonal potential creates a shear.

\newcommand{\avec}{\mathbf{a}}

Let us identify the real space edge unit vectors by the two adjacent cells $\evec_{ij} = -\evec_{ji}$ and denote the corresponding tensions as $T_{ij}$. Then the tension vectors $\hat{\Tvec}_{ij} = T_{ij} \hat{\evec}_{ij}$ form a triangulation, where $\hat{\avec}$ denotes the normal vector to $\avec$, i.e.\ $\hat{\avec}.\avec = 0$ and $||\hat{\avec}|| = ||\avec||$.

The isogonal displacement $\mathbf{u}_{ijk}$ of the real space vertices $\rvec_{ijk}$ (identified by the three adjacent cells) is given by
\begin{equation} \label{eq:isogonal_basis_vecs}
\rvec_{ijk} \to \rvec_{ijk} + \mathbf{u}_{ijk} = \rvec_{ijk} + \frac{1}{S_{ijk}} \big[ \Theta_i \Tvec_{jk} + \text{(cyc.)} \big]
\end{equation}
where $S_{ijk} = \hat{\Tvec}_{ij}.\Tvec_{ik}$ is the area of the tension triangle $(ijk)$.

First, observe that the uniform isogonal mode $\Theta_i = {}$const.\ has no effect on the vertex positions because
\begin{equation}
    \Tvec_{jk} + \Tvec_{ki} + \Tvec_{jk} = 0 \quad \text{(force balance).}
\end{equation}
Now we aim to show that a constant gradient in $\Theta_i$ drives a uniform displacement of the $\rvec_{ijk}$. Specifically, by uniform gradient, we mean $\Theta_i = \tvec_i.\avec$, i.e.\ a linear gradient in the tension space ($\tvec_i$ is the position of the tension triangulation vertex corresponding to cell $i$, such that $\hat{\Tvec}_{ij} = \tvec_j - \tvec_i$).
To show that the displacement in real space is uniform, it is enough to show that two adjacent vertices are displaced identically. By induction, this implies that all displacements are identical. It is therefore sufficient to consider a quartet of cells ($i = 1{-}4$), corresponding to a ``kite'' in tension space (note that $\hat{\mathbf{a}}.\mathbf{b}$ is identical to the wedge product $\mathbf{a} \wedge \mathbf{b}$.)

Because a constant can be arbitrarily added to all $\Theta_i$, we can set $\Theta_1 = 0$ and thus have $\Theta_i = \hat{\Tvec}_{ij}.\avec$ for $i = 2,3,4$. The displacements now read
\begin{align}
    \mathbf{u}_{123} &= \frac{1}{S_{123}} \big(
    \hat{\Tvec}_{12}.\avec \; \Tvec_{31} +
    \hat{\Tvec}_{13}.\avec \; \Tvec_{12}
    \big) \label{eq:u123} \\
    \mathbf{u}_{134} &= \frac{1}{S_{134}} \big(
    \hat{\Tvec}_{13}.\avec \; \Tvec_{41} +
    \hat{\Tvec}_{14}.\avec \; \Tvec_{13} \big)
    \label{eq:u134}
\end{align}
To show that these displacements are identical, we project them onto two conveniently chosen, linearly independent vectors, namely $\hat{\Tvec}_{12}$ and $\hat{\Tvec}_{13}$.
For the latter, we find
\begin{align}
    \hat{\Tvec}_{13}.\mathbf{u}_{123} &= \frac{1}{S_{123}} \hat{\Tvec}_{13}.\avec \; \hat{\Tvec}_{13}.\Tvec_{12} = -\hat{\Tvec}_{13}.\avec, \\
    \hat{\Tvec}_{13}.\mathbf{u}_{134} &= \frac{1}{S_{134}} -\hat{\Tvec}_{13}.\avec \; \hat{\Tvec}_{13}.\Tvec_{14} = -\hat{\Tvec}_{13}.\avec,
\end{align}
where we used that $\hat{\Tvec}_{ij}.\Tvec_{ij} = 0$ and applied the definition of $S_{ijk}$.

Projecting \eqref{eq:u123} and \eqref{eq:u134} onto $\hat{\Tvec}_{12}$ gives
\begin{align}
    \hat{\Tvec}_{12}.\mathbf{u}_{123} &= \frac{1}{S_{123}} \hat{\Tvec}_{12}.\avec \; \hat{\Tvec}_{12}.\Tvec_{31} = -\hat{\Tvec}_{12}.\avec  \label{eq:T12.u123}\\
    \hat{\Tvec}_{12}.\mathbf{u}_{134} &= \frac{1}{S_{134}} \big( \hat{\Tvec}_{13}.\avec \; \hat{\Tvec}_{12}.\Tvec_{41} + \hat{\Tvec}_{14}.\avec \; \hat{\Tvec}_{12}.\Tvec_{13} \big) \label{eq:T12.u134}
\end{align}
To show the equality of these two right-hand sides, we use that given $\hat{\Tvec}_{13}.\avec$ and $\hat{\Tvec}_{14}.\avec$ we can find $\avec$ and substitute the result into $\hat{\Tvec}_{12}.\avec$. 
We start by ``expanding the identity''
\begin{equation}
    \begingroup \renewcommand*{\arraystretch}{1.2}
    \begin{pmatrix}
        \text{---}  \hat{\Tvec}_{13} \text{---}  \\ \text{---}  \hat{\Tvec}_{14} \text{---} 
    \end{pmatrix}\avec
    =
    \begin{pmatrix}
        \hat{\Tvec}_{13}.\avec \\ \hat{\Tvec}_{14}.\avec
    \end{pmatrix}
    \quad \Rightarrow \quad
    \avec = 
    \begin{pmatrix}
        \text{---} \hat{\Tvec}_{13} \text{---}  \\ \text{---}  \hat{\Tvec}_{14} \text{---} 
    \end{pmatrix}^{-1}
    \begin{pmatrix}
        \hat{\Tvec}_{13}.\avec \\ \hat{\Tvec}_{14}.\avec
    \end{pmatrix}
    \endgroup
\end{equation}
Explicitly writing out the inverse matrix then gives
\begin{equation}
    \avec = \frac{1}{\hat{\Tvec}_{13}.\Tvec_{14}} \begin{pmatrix}
        \vert & \vert \\
        \Tvec_{14} & -\Tvec_{13} \\
        \vert & \vert
    \end{pmatrix}
    \begingroup \renewcommand*{\arraystretch}{1.2}
    \begin{pmatrix}
        \hat{\Tvec}_{13}.\avec \\ \hat{\Tvec}_{14}.\avec
    \end{pmatrix}
    \endgroup
\end{equation}
With this, we find the relation
\begin{equation}
    \hat{\Tvec}_{12}.\avec = -\frac{1}{S_{134}} \big( 
        \hat{\Tvec}_{12}.\Tvec_{41} \; \hat{\Tvec}_{13}.\avec
        +
        \hat{\Tvec}_{12}.\Tvec_{13} \; \hat{\Tvec}_{14}.\avec
    \big)
\end{equation}
where we used $\Tvec_{ij} = - \Tvec_{jk}$ to flip the indices on $\Tvec_{14}$. Comparing to \eqref{eq:T12.u123} and \eqref{eq:T12.u134} now shows the identity of their RHSs. 

Taken together, we have shown that 
\begin{equation}
    \hat{\Tvec}_{12}.\mathbf{u}_{123} = \hat{\Tvec}_{12}.\mathbf{u}_{134}
    \quad \text{and} \quad
    \hat{\Tvec}_{13}.\mathbf{u}_{123} = \hat{\Tvec}_{13}.\mathbf{u}_{134},
\end{equation}
Because $\hat{\Tvec}_{12}$ and $\hat{\Tvec}_{13}$ are linearly independent, it follows that $\mathbf{u}_{123} = \mathbf{u}_{134}$. QED.

We just showed that a constant gradient in $\Theta_i$ corresponds to a uniform displacement of the real space vertices $\rvec_{ijk}$. 
We can therefore think of $\Theta(\tvec)$ as a ``potential'' for the isogonal displacement field: $\mathbf{u} \approx \nabla_{\tvec} \Theta$, where the approximation is valid for slowly varying gradients and exact for constant gradients. The gradient $\nabla_{\tvec}$ is taken in \emph{tension space} because the function $\Theta(\tvec_i) = \Theta_i$ is defined on the vertices of the tension triangulation $\tvec_i$.

A pure shear aligned with the coordinate axes is given by a displacement field $\mathbf{u}(\rvec) = \varepsilon \; \mathrm{diag} \, (1, -1).\rvec$ and is therefore generated (approximately) by a quadratic isogonal potential $\Theta_i = \varepsilon \; \tvec_i^\mathrm{T}.\mathrm{diag} \, (-1, 1).\tvec = \varepsilon \, [(t_i^1)^2 - (t_i^2)^2]$.

\subsection{Isogonal Hessian and tissue-scale isogonal shear modulus}

Above we have shown that isogonal modes can create a shear (on the level of cell centroids). We now show (a) that shear transformations are the lowest-energy isogonal modes and hence will dominate the response of the tissue to external forces, and (b) that shear transformations incur a finite energy cost, implying a finite shear modulus -- the tissue is not a fluid.
To do this, we assume that the tissue is initially at the elastic energy minimum under the constraints of a given tension tension triangulation. In this case, we can approximate the energy $E$ by a quadratic function of displacement by computing the Hessian. In our model we impose that vertex angles are given by the intrinsic tension triangulation, hence we cannot study how tensions respond to external forces. Further, due to the dominance of cortical tensions, the lowest-energy deformations of the tissue are isogonal.
Hence, here, we only allow isogonal displacements, and we will project the Hessian onto the space of isogonal modes. The lowest eigenvectors of the isogonal Hessian will then correspond to the modes excited by external forcing.
Note that isogonal deformations only leave the dominant junctional elastic energy $\sum_{ij} T_{ij}\ell_{ij}-p\sum_i A_i$ invariant if they don't deform the tissue boundary.
The tissue patch (shown in Fig.~\ref{SI-fig:isogonal-shear}A) we consider in the following calculation should therefore be imagined as embedded into a larger cell array with overall fixed boundaries, as in the simulations of Fig.~\ref{fig:tissue-model-mutants} where the net shear of active and passive regions is zero.

The isogonal unit vectors convert the isogonal potentials into real-space displacement of vertices: the isogonal potential $\Theta_j$ at cell $j$ contributes a term 
$\Theta_j \mathbf{I}_j^{jkl}$ to the displacement of vertex $\mathbf{r}_{jkl}$ adjacent to $j$. The $\mathbf{I}_j^{jkl}$ are defined by the action of isogonal modes \eqref{eq:isogonal_basis_vecs} and serve to project the Hessian onto the space of isogonal modes. Explicitly, $I_{j}^{jkl} = \tfrac{1}{S_{jkl}} \Tvec_{kl}$, where $S_{jkl} = \hat{\Tvec}_{jk}.\Tvec_{kl}$ is the area of the tension triangle $(jkl)$. The isogonal Hessian reads:
\begin{align}
    \mathcal{H}_{il} = \sum_{(jk),(mn)} \: \mathbf{I}_i^{ijk}\frac{\partial^2 E(\{\mathbf{r}_{abc} \}) }{\partial \mathbf{r}_{ijk} \partial \mathbf{r}_{lmn}} \mathbf{I}_l^{lmn}
\end{align}
Note that only the $E_\mathcal{C}$ term from \eqref{eq:total-energy} contributes to the isogonal Hessian, by virtue of the isogonal projection.

With the same code used for the tissue-scale simulations, we can numerically construct and diagonalize $\mathcal{H}_{il}$. There are 3 trivial 0-modes which correspond to two real-space translations and a globally constant isogonal parameter, which creates no cell displacement. This is in accordance with the results from above and the fact that the $E$ only depends on differences in vertex positions. We find that the two lowest non-trivial modes are shear modes (quadratically varying isogonal potential $\Theta_i$). They are shown for an example lattice in Fig.~\ref{SI-fig:isogonal-shear}A. Finding shear modes is not unexpected since they minimize the variation in vertex displacement across the tissue, and the energy function $E$ is convex (quadratic).

External forces, e.g. external shear, will mainly excite the lowest eigenmodes of the Hessian. The energy cost is determined by the Hessian eigenvalues, which therefore define the shear modulus. In the example shown in Fig.~\ref{SI-fig:isogonal-shear}A the eigenvalues of the shear modes are 2 orders of magnitude smaller than the next eigenvalues. The highest eigenvalues correspond to isogonal deformations localized to a single quartet.

To compute a measure of resistance to shear deformations, the tissue-scale ``isogonal shear modulus'', we consider a vertex displacement $h\cdot \delta \mathbf{r}_{ijk}, \; h>0$ along one of the two lowest-energy shear eigenmodes of the isogonal Hessian. We then compute
\begin{align}
    \label{eq:isogonal_shear_modulus}
    \mu_\mathrm{tissue} = \frac{\partial^2}{\partial s(h)^2} \sum_l E_{\mathcal{C},l}(\{\mathbf{r}_{ijk}+ h\delta \mathbf{r}_{ijk} \} )
\end{align}
where $s(h)$ is the shear induced by the vertex displacement. In practice, we compute $s(h)$  by least-squares fit of a shear transformation to the displacement map $\mathbf{r}_{ijk}\mapsto  \mathbf{r}_{ijk}+h\cdot\delta \mathbf{r}_{ijk}$). We then fit a parabola to the curve $(s(h), E_\mathcal{C}(h))$ to extract the $2^\mathrm{nd}$ derivative.
We can then investigate how the isogonal shear modulus depends on the cell-level shear modulus in the elastic energy. As expected, we find that the two are proportional (Fig.~\ref{SI-fig:isogonal-shear}B), showing that the tissue as a whole is solid, in the sense of a non-vanishing shear modulus.
We note that a complete investigation of the tissue-scale shear modulus, i.e.\ a complete account of tissue rheology, will require addressing (a) energy costs due to deformations of the tissue boundary and (b) the question of how tensions change when external forces are applied and is hence beyond the scope of the paper.

\section{Model details and simulation methods}
\label{SI:simulation_methods}

In the following, we explain the simulations presented in the paper and provide the modeling and implementation details. These comprise the mean-field calculation of the dynamics of the distribution of vertex angles, the symmetric lattice of a single intercalating quartet (shown in the companion paper \cite{Brauns.etal2024elife}), and the tissue-scale simulation of a disordered epithelial tissue comprising both active and passive regions. All code used to run and analyze the simulations is available on GitHub \url{https://github.com/nikolas-claussen/CE\_simulation\_public}. 

\subsection{Positive feedback tension dynamics and single-vertex simulation}

We begin by describing the intrinsic tension dynamics of a single tension triangle. This is the basis for the singe-triangle simulation in Figs.~\ref{fig:tension-config-data}B and~\ref{fig:tension-config-data}C, where we show the dynamics of the distribution of vertex angles predicted by a positive tension feedback model. Under this model, the edge tensions $\tilde{T}_\alpha$, $\alpha = 1{-}3$, at a vertex evolve according to:
\begin{align*}
    \tau_\mathrm{T}^{} \dot{\tilde{T}}_\alpha = T_\alpha^n - \frac{1}{3}\sum_\beta \tilde{T}_\beta^n
\end{align*}
where $n>1$ and $\tau_\mathrm{T}^{}$ is a time scale converting simulation time into minutes. Note that we fix the overall tension scale by the constraint that the triangle perimeter $P=\sum_\alpha \tilde{T}_\alpha$ is fixed (corresponding to a finite total myosin pool). This simplified the analysis of saturating positive feedback below and is consistent with the single-quartet simulations in the companion paper. Fixing the triangle area $A$ instead does not significantly affect the results for the single triangle simulations. 
The feedback exponent was taken as $n=4$ as in Ref.~\cite{Brauns.etal2024elife}. Different values give very similar results, so no systematic optimization was performed.
We note that since the tension constraint $P=\mathrm{const.}$ or $A=\mathrm{const.}$ only affects the overall triangle size, the trajectories in LTC space (but not their speed) are independent of it.

We integrated the tension dynamics using a 4th order Runge--Kutta method as implemented in the \texttt{SciPy} software package \cite{Virtanenetal.2020}, using as initial conditions the vertex angles in the experimental data at time $t=5\:\text{min}$ (the vertex angles from the data were temporally smoothed with a window of $2\:\text{min}$ to reduce noise). From our simulation, we computed the LTC parameter to compare the marginals with the experimental data. 

\paragraph*{Dynamics in the LTC space.}
To find the trajectories in the LTC space, we integrated \eqref{eq:tension-dynamics}. A change of variables from side length to the LTC space yields the flow shown in Fig.~\ref{fig:tension-config-data}C. To model the effect of intercalations, which remove highly obtuse tension triangles from the distribution, we stopped integration once the maximum relative tension increased over $T_\mathrm{rel}=1.56$, the median of the relative tension at the moment of a T1 observed the tissue-scale simulations with isogonal stretching (as well as in the experimental data). These ``collapse'' triangles were removed from the simulation ensemble.

\paragraph*{Fit of the tension timescale to experimental data.}

To determine the tension timescale $\tau_T$, we compute the dynamics of the relative tension (the ratio of the edge tension to that of the adjacent edges) for all edges that undergo a T1 
transition and align them temporally based on the time the T1 process occurs. This leads an average $T_\mathrm{rel}(t)$ that can be compared to the time traces from the experimental data reported in the companion paper\cite{Brauns.etal2024elife} to fit $\tau_T$.
Because the tension dynamics in the tissue-scale simulations are affected by tension balancing (see below), while the single-triangle and single-quartet simulations are not, we fit $\tau_T$ separately for the two classes of simulations. 
Note that we fit the tension timescale to a microscopic process, the T1 transition, and the resulting time course for macroscopic observables (tissue extension, LTC distribution) matches the experimental data without additional adjustment.

\subsection{Tension dynamics post intercalation: Myosin handover and passive tension}

In the single-triangle simulations, we only consider dynamics up to the moment of intercalation. At this point, the tension triangulation is modified topologically: the edge corresponding to the collapsed interfaces is replaced by one corresponding to the new interface (triangulation ``flip''). To complete our model, we must specify the initial conditions of this new edge.

As explained in the companion paper, we propose a myosin handover mechanism to explain the extension of the new interface post-intercalation. An interface with cortical tension $T$ is comprised of the adherens-junctional actomyosin cortex of the two adjacent cells, which are coupled mechanically via adherens junctions. Under force balance, the total tension $T$ has to be constant along the cortex, but the individual tensions on either side can be non-uniform, as the resulting traction forces are exchanged via adherens junctions. In the following, we assume as a first-order approximation that the level of active tension (i.e.\ myosin concentration) varies linearly along an interface (similar calculations have been performed in Ref.~\cite{Kale.etal2018} to calculate interfacial shear stress). This allows us to geometrically obtain the myosin concentration at the individual cortices that will form the two juxtaposed sides of the new interface (see companion paper).

Consider a vertex, where interfaces $i=0, 1, 2$ (with tensions $T_0, T_1, T_2$) meet, and let $0$ be the interface about to collapse. For an illustration, see Ref.~\cite{Brauns.etal2024elife}, Fig.~4). The three cells that meet at the vertex will be referred to by $(01), (12), (21)$ (where e.g. $(01)$ is the cell abutting interfaces $0$ and $1$). Let $m_{01}, m_{12}, m_{21}$ be the motor molecule concentrations (in units of tension) at the vertex in the junctional cortices of the three cells. Then, the tensions are related to the motor molecule concentrations as
\begin{align*}
    T_0 = m_{21}+m_{01}, \quad
    T_1 = m_{01}+m_{12}, \quad
    T_2 = m_{12}+m_{21}.
\end{align*}
This uses the assumption of myosin continuity at vertices and the fact that the tension on an interface is the sum of the tensions of the two cortices that make it up. The motor molecule concentration on the cortex belonging to the new interfaces post-collapse will be equal to $m_{12}$. Solving for this in terms of the tensions:
\begin{align*}
    m_{12} = \frac{T_1+T_2-T_0}{2}
\end{align*}
The new interface consists of two cortices, coming from the two vertices of the collapsed interface. Let the tensions at the two triangles be $T_0, T_1, T_2$ and $T_0, T_1', T_2'$. Let $T_{a}$ be the active tension on the new interface immediately after the T1. It is equal to
\begin{align} \label{eq:passive-tension-initial}
    T_{a} &=  \frac{(T_1+T_2-T_0)+(T_1'+T_2'-T_0)}{2}
\end{align}
Note however that the total tension $T_n$ on the new interface is not necessarily equal to $T_{a}$. The total tension is defined geometrically from the angles at the new interface (or, equivalently, the tension triangulation vertices). Indeed, generally, $T_n>T_a$, i.e.\ the active tension on the new interface is not enough to balance the tension due to the adjacent edges. As explained in the companion paper, we introduce a passive tension $T_p$ on the new edge which balances this deficit
\begin{align*}
    T_{p} = T_{n} -  T_{a} = T_{n} -  (m_{12}+m_{12}')
\end{align*}
For example, if a perfectly symmetric quartet collapses when the vertex angle facing the collapsing edge is $90^\circ$, $T_1=T_2=T_1'=T_2'=1$ and $T_0=T_n=\sqrt{2}$. Therefore, $T_{a,n}=2-\sqrt{2}\approx 0.6$ and $T_{p, n} =\sqrt{2}-(2-\sqrt{2}) \approx 0.8$. Note that by the triangle inequality, for any convex quadrilateral with perimeter $P$ and diagonals $D_1, D_2$, one has $P/2 \leq D_1+D_2 \leq P$. Applying this to the quadrilateral formed by the two tension triangles at the collapsing interface, we get $T_{a,n} \geq 0$ and $T_{p,n}\geq 0$: the handover formula always results in positive active and passive tensions. Further, the ``handover'' mechanism robustly generates irreversible T1s: if an interface were to collapse back after a T1, the newly formed interface would inherit high myosin levels and therefore be likely to collapse again.

The passive tension subsequently relaxes visco-elastically with rate $\tau_p^{-1}$:
$\dot{T_p} = -\tau_p^{-1} T_p$. Combining this with the feedback equation, the evolution of the total tension is 
\begin{align}
    \dot{\tilde{T}} = \tau_T^{-1} \left( T^n - \frac{1}{3}\sum_\beta \tilde{T}_\beta^n \right) - \tau_p^{-1} T_p
\end{align}
where we assumed that the positive feedback only operates on the active tension. The relaxation time scale was chosen at $\tau_p=\tau_T/4$, which fits the observed decay of tension post T1. If the tension relaxation rate $\tau_p^{-1}$ is too low, then positive tension feedback leads to a reversal of active T1s, and the T1 gets stuck at the 4-fold vertex configuration.

\subsection{Cell shape elasticity} 

As explained in the main text, in our model,  tension dynamics translate into tissue dynamics via force balance which we implement by minimizing the energy function \eqref{eq:total-energy}. Here, we explain our choice of the term representing cell-shape elasticity, $E_C$. We use an elastic energy based on the cell shape tensor
\begin{align*}
   S_\mathcal{C} = \sum_i \frac{\mathbf{e}_i \otimes \mathbf{e}_i}{|\mathbf{e}_i|}
   \label{eq_cell_shape}
\end{align*}
where $i$ runs over the edges of the cell $\mathcal{C}$, and $\mathbf{e}_i$ is the vector pointing from one vertex of the interface to the other. Note that in this shape tensor, each edge contributes linearly in length to the cell shape. This means that artificially subdividing an edge has no effect on the cell shape tensor. This makes sense if we assume that the elasticity we aim to model using $S$ resides in the cell interior (incompressibility, microtubules, intermediate filaments, nucleus). The shape tensor can also be defined using vectors from the cell centroid to its vertices (in the lattice, the two definitions are equivalent). An alternative definition of the elastic energy using $\tilde{S}_\mathcal{C} = \sum_i \mathbf{e}_i \otimes \mathbf{e}_i$ instead gives broadly similar results (although interface collapse happens more abruptly, because of the higher order non-linearity).

We assign a reference shape $S_0$ and define the cell elastic energy via 
\begin{align}
    E_\mathcal{C} = \lambda [\mathrm{Tr}(S_\mathcal{C}-S_0)]^2 + \mu \mathrm{Tr}[(S_\mathcal{C}-S_0)^2]
\end{align}
with bulk modulus $\lambda$ and shear modulus $\mu$. The reference shape $S_0$ controls the isogonal mode. An isotropic $S_0 \propto \mathrm{Id}$ favours equilateral hexagons.
We used a shear/bulk ratio of $\mu/\lambda=1$ for all ``active'' cells in the simulations. Because of the separation of scales between cortical tensions and elastic energy baked into the model, the absolute values of $\lambda, \mu$ are irrelevant. 

Strikingly, when the cell array is a regular lattice (see Fig.~\ref{fig:intro}G and single-quartet simulations shown in the companion paper \cite{Brauns.etal2024elife}), the ``shape strain'' $S_\mathcal{C}-S_0$ can always be relaxed to 0 by choice of the edge lengths, and the energy $E_\mathcal{C}=0$ throughout. This can already be seen from a degree-of-freedom count (three $\ell_i$ for the three independent components of $S_\mathcal{C}-S_0$).
The elastic energy therefore acts only on the isogonal modes (i.e.\ $\partial_{\phi_i} E_\mathcal{C} |_{\ell_i=\mathrm{minimizers}} =0$) and there is no energy barrier for intercalations. Consequently, there is no need for noise to drive intercalations in our model. By contrast, for the widely-used area-perimeter elastic energy $E=(A-A_0)^2+(P-P_0)^2$ (where $A,P$ are the cell area and perimeter, and $A_0, P_0$ their target values) \cite{Bi.etal2015}, there exists an energy barrier, and the inner interface $\ell_0$ only collapses when $\beta \to \pi$ (see Fig.~\ref{SI-fig:quartet-model-energy-barrier}B). Note that the area-perimeter energy is a special case because of the geometric incompatibility of area and perimeter constraints when $P_0/\sqrt{A_0} < 3.72$.
For $P_0/\sqrt{A_0} > 3.72$ cell shape is underdetermined (floppy) and the shear modulus vanishes. Recall that we have shown that a finite shear modulus is required to translate tension dynamics into tissue-level convergent extension (see Fig.~\ref{fig:tissue-model-mutants}F).
The area-perimeter energy in the fluid regime is therefore not suitable to model convergent extension driven by internal tension dynamics.

Combining area elasticity with shear elasticity based on the shape tensor, $E=(A-A_0)^2 + \mu \mathrm{Tr}[(S_\mathcal{C}-S_0)^2]$, (or perimeter elasticity with cell shape bulk elasticity) leads to similar results as \eqref{eq:cell_shape_energy}.
Note also that because of the degree-of-freedom count, the system is under-specified if the shear modulus is 0, foreshadowing the fact that without shear modulus, no convergence-extension takes place. 
Indeed, in the case of the symmetric lattice, the degrees of freedom are the three vertex angles, determined by the tensions, and the three interface lengths, independent of the angles. Fixing the cell area still leaves freedom to change the interface lengths. More generally, we have shown above and illustrated in Fig.~\ref{fig:intro} that isogonal modes allow for area-preserving shear transformations.

\subsection{Tissue scale simulation}

Using the components of the single-quartet simulation, we now describe a simulation of an arbitrary, disordered tissue patch. The key assumption is again adiabatic force balance and cortical tension dominance: the tissue geometry is found by solving for force balance, by first finding the angles from the cortical tensions and then the isogonal mode by minimizing cell elastic energy. The key quantity is the tension triangulation whose dynamics determine the tissue dynamics.

Conceptually, there is one main novel element: since the tension triangulation is now disordered and comprises many heterogeneous tension triangles, an additional mechanism is required to ensure the tension triangles fit together to a global triangulation. Further, the higher computational load requires different numerical methods.

\paragraph*{Data structure.}

We model a piece of epithelial tissue by a polygonal tiling of the plane, polygons corresponding to cells. Two cells are neighbors if they share an edge (this implies that the cell adjacency graph is a triangulation). 

Following the idea of the dual tension triangulation introduced in the main text, a polygonal tiling is represented by a triangular mesh, in which cells are represented by vertices, cell edges by triangulation edges, and cell vertices by triangulation faces. Vertices, edges, and faces have the following attributes which are used for time evolution:
\begin{itemize}
    \item Vertices: dual coordinates, the $x,y$ coordinates of the vertex in the tension triangulation which will be denoted $\mathbf{t}_i$, and the reference shape $g_0$ for the shape-tensor elastic energy.  
    \item Edges: the active and passive tensions on an edge.
    \item Faces: primal coordinates, the $x,y$ coordinates of the vertex in the cell tessellation corresponding to the face, denoted by $\mathbf{r}_{ijk}$.
\end{itemize}
Given the dual coordinates of two adjacent vertices $\mathbf{t}_i,\mathbf{t}_j$, the tension on the edge connecting cells $i,j$ is $T_{ij}=|\mathbf{t}_i-\mathbf{t}_j|$. The dual coordinates encode a configuration of balanced cortical tensions. Note that the primal (cell) and dual (tension) coordinates live in different spaces, and the overall scale of the tension coordinates is irrelevant.

The mesh is implemented in object-oriented python, using a half-edge mesh data structure \cite{Marschner.Shirley2018}. A half-edge mesh consists of vertices, faces, and, for each edge, two oriented half-edges. A half-edge stores a reference to the \texttt{next} half-edge on the counter-clockwise oriented face it belongs to, and a reference to its \texttt{twin}, the half-edge with opposite orientation on the adjacent face. This structure defines an orientation for every triangular face and makes mesh traversal (e.g. get all vertices of a cell) and mesh modification (e.g. intercalations) very convenient. 

\paragraph*{Overview.}

Due to the assumption of adiabatic force balance, the tissue geometry is determined by the instantaneous tensions. The dynamics are therefore determined by the change in the tension triangulation (due to positive feedback). In detail, the time evolution in our simulation is done in three steps:
\begin{enumerate}
    \item Local Euler step: For each triangular face, update the active tensions according to positive tension feedback, and the passive tensions and the rest shape according to viscous relaxation.
    \item Triangulation flattening: Optimize the dual vertex positions so that the length of each half-edge is as close as possible to its internally stored tension. This ensures global tension balance. 
    \item Cell shape optimization: Optimize the primal vertex positions to minimize the shape-tensor-based elastic energy, while constraining the angles by the dual triangulation (``isogonal constraint'').
    \item Topological modifications: Check if any primal edge has collapsed and carry out intercalations if necessary, re-initializing the active and passive tensions on the new edge.
\end{enumerate}
The next sections describe them in detail. Fig.~\ref{SI-fig:simulation_flow_chart} provides a summary of the simulation workflow.
The most complicated and time-consuming of them is the optimization of cell shapes under the isogonal constraint. Overall, a simulation of convergent extension of $\sim 1000$ cells takes approx. 15 minutes on a current-generation laptop.

Note that the workflow of our simulations differs somewhat from the more commonly used dissipative dynamics, in which cell geometry is updated dynamically according to the ODE $\gamma \partial_t \mathbf{r}_{ijk} = -\frac{\partial  E}{\partial \mathbf{r}_{ijk}}$. 
The dual separation of scales (fast relaxation and dominance of cortical tensions), means that a direct implementation of dissipative dynamics is numerically challenging. The elastic energy minimization can be seen as a relaxational Euler step with an adaptive step size.

Relaxation via fast dissipative dynamics is a viable approach that could indeed be used as an alternative numerical relaxation method. However, as the theory of fast variable elimination in the context of multiscale dynamics is well established \cite{VanKampen1985}, we do not expect a different scheme to affect our results to an important degree. The key ``ingredient'' of our model is not how exactly the elastic energy is relaxed, but how active tensions are regulated.

\paragraph*{Triangulation dynamics.}

For each mesh half-edge, we store the total intrinsic and passive tensions $\tilde{T}_{aij}, T_{p, ij}$. We collect the total and passive tensions from all half-edges $ij$ belonging to a triangular face $I_{ijk}=\{(ij),(jk),(ki)\}$ and carry out an explicit Euler step with step size $\sim 10^{-3}$ according to the following dynamics:
\begin{align}
    \tau_T \partial_t \tilde{T}_{ij} &= \tilde{T}_{ij}^{n} - k_\text{cutoff} \tilde{T}_{ij}^{n+1} - \frac{1}{3} \sum_{\alpha\in I_{ijk}} (\tilde{T}_{\alpha}^n-k_\text{cutoff} \tilde{T}_{\alpha}^{n+1}) - \frac{\tau_T}{\tau_p} T_{p,ij} \\
    \tau_p \partial_t T_{p, ij} &= - T_{p, ij}
    \label{eq_triangulation_euler}
\end{align}
The cutoff parameter $k_\text{cutoff}$ stops runaway positive feedback at values $T> 1/k_\text{cutoff}$. It is not strictly necessary since runaway feedback is cut short by intercalations, but helpful for long-time simulations, where in rare cases run-away tension cable configurations occur which ``blow up'' numerically. We chose $1/k_\text{cutoff} = 5\bar{T}$ where $\bar{T}$ is the overall tension scale so that it has almost no effect in normal tension configurations. We chose the feedback parameter $n=4$ and the relaxation rate $\tau_p=\tau_T/4$, the same used in the single-quartet and single-triangle simulations.

So far, the time evolution of tensions is purely autonomous (i.e.\ each tension triangle is independent of all others) and completely analogous to the single-quartet simulation. In the single-quartet simulation, all tension triangles are equal by symmetry and periodicity, and thus tile the plane as a triangulation by construction.
In general, this is no longer guaranteed. Even if one starts with a flat tension triangulation, the autonomous time evolution \eqref{eq_triangulation_euler} will result in a set of tension triangles that cannot fit together to a flat triangulation. The triangulation will develop curvature \cite{Noll.etal2020}, i.e.\ ``crumple'' out of plane.
This means that we have to add only one ingredient to enforce triangulation planarity.

We enforce planarity as follows. The set of all balanced tension triangulations is parameterized by the dual vertex positions $\mathbf{t}_i$. We, therefore, seek the balanced tension configuration that approximates the intrinsic tensions as closely as possible by minimizing the following elastic energy:
\begin{align}
    E_\text{tri} =  \frac{1}{2N_\text{edges}} \sum_{ij} (||\mathbf{t}_i-\mathbf{t}_j|| - \tilde{T}_{ij})^2 + \frac{1}{2N_\text{triangles}} \sum_{ijk}  \mathrm{Pen}(A_{ijk}, A_0)
    \label{eq_triangle_energy}
\end{align}
Since the simulation is based on half-edges, the per-edge intrinsic tension $\tilde{T}_{ij}$ is the average of the tensions of the two half-edges (which can be thought of as the two cellular cortices coupled in a cell-cell interface). The second term, $\mathrm{Pen}(A,A_0)$ is a penalty term on the triangle areas $A$ (calculated from the vertex positions $\mathbf{t}_i$). It fixes the overall tension scale (which is arbitrary due to force balance) and ensures the triangulation is well-behaved numerically. It contains two terms:
\begin{align}
    \mathrm{Pen}(A,A_0) = \gamma_\text{tri} \left( 100\cdot(\max\{0, \frac{A_0}{4}-A\})^2 + \frac{1}{2} (A-A_0)^2 \right)
\end{align}
Here, $A_0$ is an arbitrary reference area fixing the overall tension scale. Because the triangular mesh is oriented, the area $A_{ijk}$ is signed, with negative signs corresponding to un-physical ``flipped'' configurations. These, and degenerate triangles, are penalized by the first term. The second term is a soft potential fixing the overall tension scale and preventing isotropic growth. The penalty strength $\gamma$ was chosen at $10^{-3}$, so that the area penalty represents a small correction to the length-term.

In the simulation, we first carry out an explicit Euler step for the intrinsic tensions, and then ``flatten the triangulation'', i.e.\ find the dual vertex positions by minimizing \eqref{eq_triangle_energy}. This determines the extrinsic tensions $T_{ij}=||\mathbf{t}_i-\mathbf{t}_j||$. Finally, we relax the intrinsic tensions to force balance with rate $\tau_\mathrm{balance}$, 
\begin{align}
    \label{eq:balance_relax}
    \frac{d \tilde{T}_{ij}}{dt} = -\tau_\mathrm{balance}^{-1} (\tilde{T}_{ij} - T_{ij})
\end{align}
In order to ensure numerical stability even for large $\tau_T$, we use the exact solution of \eqref{eq:balance_relax} numerically, instead of carrying out an explicit Euler step.

The triangulation-flattening prescription can be thought of as modeling, indirectly, known feedback loops that lead to convergence to balanced tension, such as strain-rate feedback \cite{Noll.etal2017}, which has been demonstrated experimentally \cite{Gustafson.etal2022}. The rate $\tau_\mathrm{balance}^{-1}$ determines the speed of these feedback loops. The influence of the value of $\tau_\mathrm{balance}^{-1}$ is studied in Fig.~\ref{SI-fig:parameter_scan}. 
In principle, strain rate feedback requires movement of cell vertices to sense tension imbalance and then bring tensions back to a balanced state. We neglect this vertex motion since we assume balance is restored rapidly.
Directly modeling of strain rate feedback requires going beyond the separation between tension- and real-space-dynamics that is the basis of our model here and is therefore beyond the scope of this paper. Further, the triangulation-flattening prescription is the most parsimonious generalization of the local tension feedback dynamics employed in the single-quartet simulation to a disordered tissue patch.

The complete tension dynamics that are implemented in the simulation therefore comprises three terms in addition to the dynamics shown in the main text \eqref{eq:tension-dynamics}: (1) Numerical cutoff of positive feedback at very high tensions (2) Passive tension dynamics post T1 (3) Relaxation of tensions to force equilibrium:
\begin{align}
    \partial_t \tilde{T}_{ij} = &\tau_\mathrm{T}^{-1} \left( \tilde{T}_{ij}^{n} - k_\text{cutoff} \tilde{T}_{ij}^{n+1} - \frac{1}{3} \sum_{\alpha\in I_{ijk}} (\tilde{T}_{\alpha}^n-k_\text{cutoff} \tilde{T}_{\alpha}^{n+1}) \right) \\
    &- \tau_p^{-1} T_{p, ij} - \tau_\mathrm{balance}^{-1} (\tilde{T}_{ij} - T_{ij})
\end{align}
We find that our simulations are robust to different choices of the tension dynamics parameters $\tau_p, \tau_\mathrm{balance}, k_\mathrm{cutoff}$, as demonstrated in Fig.~\ref{SI-fig:parameter_scan}. We can, for instance, decrease the balancing time by $40\times$ without affecting the qualitative behavior.

\paragraph*{Cell shape dynamics.}

Above, we have described the dynamics of the tension triangulation. After an Euler step to update the tensions has been taken, we calculate the real space cell tessellation (i.e.\ the positions of the cell vertices) by minimizing the cell-shape elastic energy while constraining the angles by the tension triangulation. In the case of a single quartet of identical cells, we could explicitly parameterize the cell shape by interface lengths and angles, and optimize only with respect to the former. In the disordered case, it is more convenient to enforce the angle constraint via a (strong) energy penalty. It would also be possible to instead directly optimize the per-cell isogonal potentials $\Theta_i$ with respect to the Voronoi reference, obviating the need for a penalty.

The elastic energy to minimize therefore has two terms -- the cell elastic energy \eqref{eq:cell_shape_energy} calculated from cell shape tensor $S_\mathcal{C}$ \eqref{eq_cell_shape} and reference shape $S_0$, and the angle penalty, which corresponds to the cortical tension energy term:
\begin{align}
    E = \frac{1}{2N_\text{edges}} \sum_{ij}  T_{ij} \overline{\ell} (1- \hat{n}_{ij} \cdot \hat{\mathbf{r}}_{ij})
      + \frac{\varepsilon}{N_\text{cells}}\sum_i E_{\mathcal{C},i}
    \label{eq_energy_angle_penalty}
\end{align}
Here, $\hat{n}_{ij}$ is the unit normal of the tension edge, $\hat{\mathbf{r}}_{ij}$ the unit tangent vector of the primal edge, and
$\overline{\ell}$ is the average interface length, to ensure the term has the correct dimensions of energy.
The small parameter $\varepsilon$ (we use $\varepsilon \sim 10^{-3}\lambda^{-1}$) represents the fact that the cell shape energy is assumed to be much weaker than the cortical tensions.
If desired, one can replace or modify the cell shape elastic energy $E_\mathcal{C}$ in \eqref{eq_energy_angle_penalty} by another energy, e.g. the common area-perimeter elasticity.

In \eqref{eq_energy_angle_penalty}, we use a different term than $T_{ij} \ell_{ij}$ to represent angle constraint due to cortical tensions.
When determining the vertex positions $\mathbf{r}_{ijk}$, we minimize the cell shape elastic energy under the constraint of fixed angles, as given by the tension triangulation (so in a sense, we minimize the energy not with respect to the complete vertex positions, but with respect to the isogonal modes only). The tensions are externally imposed parameters and not functions of the vertex positions, like in a spring network. \eqref{eq_energy_angle_penalty} does not represent the physical energy, but merely the way we numerically enforce the angle constraints.
We use the alternative energy term $T_{ij}(1- \hat{n}_{ij} \cdot \hat{\mathbf{r}}_{ij})$ for two numerical reasons. (1) Compared to $T_{ij} \ell_{ij}$, it does not require a pressure term $p \sum_i A_i$ to avoid global constriction (2) it avoids potential convergence issues when minimizing the elastic energies in the presence of very short junctions, which occur before and after T1-transitions (3) it automatically penalizes invalid self-intersecting vertex configurations. However, if the boundary is fixed, both forms of the energy lead to the same family of ground states, namely those where the vertex angles are complementary to those of the tension triangulation. The energy term $T_{ij}(1- \hat{n}_{ij} \cdot \hat{\mathbf{r}}_{ij})$ was explicitly constructed to ensure this. Away from the ground state, the two energies differ; however, this is immaterial since we work in the regime where the tension constraints are always satisfied exactly.
We can also rewrite
\begin{align}
    T_{ij} (1- \hat{n}_{ij} \cdot \hat{\mathbf{r}}_{ij}) = \frac{T_{ij}}{2} ||\hat{\mathbf{r}}_{ij}-\hat{n}_{ij}||^2 = \frac{1}{2} ||\mathbf{T}_{ij} - R(\tfrac{\pi}{2})\mathbf{t}_{ij}||^2
\end{align}
where $\mathbf{T}_{ij}$ is the real-space tension vector, $R(\tfrac{\pi}{2})$ a $90^\circ$ rotation matrix, and $\mathbf{t}_{ij}=\mathbf{t}_i-\mathbf{t}_j$ the dual tension vector, defined by the tension vertices. This shows that minimization of the $\sum_{ij} T_{ij}(1- \hat{n}_{ij} \cdot \hat{\mathbf{r}}_{ij})$ energy enforces that the tensions in the cell array match those prescribed by the tension triangulation.

After carrying out a local Euler step and updating the dual vertex positions by ``triangulation flattening'', we find the primal vertex positions by minimizing the elastic energy \eqref{eq_energy_angle_penalty}.

\paragraph*{Topological modifications.}

After each time step, we check whether any edge has collapsed. We call an edge collapsed if its length is $<10\%$ of the overall length scale (mean edge length at simulation start). If this occurs, we carry out a neighbor exchange (flip the corresponding edge in the triangulation) and set the active and passive tensions on the new edge as described above. 

Edges that underwent a T1 transition in the last 20 simulation time steps (corresponding to 2.5 minutes) are excluded from edge collapse to prevent the immediate re-collapse of newly formed junctions. Instead of this hard-coded criterion, one could also calculate the strain rate of junctions of short length and exclude extending junctions, as shown by the time traces of junction length in Fig.~\ref{SI-fig:cauterized}B.

\paragraph*{Automatic differentiation and numerical optimization.}

The simulation requires solving two optimization problems, one for the triangulation \eqref{eq_triangle_energy} and one for the real-space shape \eqref{eq_energy_angle_penalty}. To solve an optimization problem efficiently, it is crucial to know the gradient of the objective function. We use the python library \texttt{JAX} \cite{Bradbury.etal2018} to automatically differentiate our energy functions, allowing easy, rapid, and bug-free calculation of the Jacobians. We further use \texttt{JAX}'s just-in-time functionality to accelerate our Python code. Gradient-based optimization requires the objective function to be at least once differentiable. To ensure this, we mollify functions as required; for example, the Euclidean norm $\sqrt{x^2+y^2} \rightarrow \sqrt{x^2+y^2+\varepsilon^2}, \: \varepsilon\sim 10^{-3}$ to ensure differentiability when computing the lengths of very short edges.

With these tools, the energies can be efficiently minimized using \texttt{scipy}'s conjugate-gradient optimizer. With a relative tolerance of $10^{-5}$, a shape optimization step for $\sim 1000$ cells takes $\sim 10\text{s}$ on a laptop. Using the half-edge data structure described above, and \texttt{JAX}-based automatic differentiation, we can also easily construct the isogonal Hessian $\mathcal{H}_{ij}$ analyzed in Fig.~\ref{SI-fig:isogonal-shear}.

\paragraph*{Creation of initial conditions.}

Finally, we need to specify the simulation the initial conditions. Note that because the cell vertex positions are computed from the tension triangulation, in our initial condition we specify only the tension triangulation.

We consider two types of initial conditions, using either a regular lattice or a random point process to choose the tension vertices. In the former case, we chose rectangular patches of a regular triangular lattice, with i.i.d. Gaussian noise added to the vertex positions. The amplitude of the noise was chosen so that $\mathrm{std}(T_{ij})=0.1\:\mathrm{mean}(T_{ij})$.
The influence of the Gaussian noise standard deviation is shown in Fig.~\ref{SI-fig:parameter_scan}.
A triangular lattice corresponds to a hexagonal lattice of cells. Indeed, locally, the \textit{Drosophila} embryo at the onset of GBE looks like it is composed of (noisy) hexagonal lattice patches, as shown in Fig.~\ref{SI-fig:hexagonal-order}. This lattice is subjected to a small initial strain to create an initial tension anisotropy, as described in the main text.

To investigate the effect of disorder, we also create initial conditions using a hard disk process. This randomly samples 2d-points (the eventual vertices in tension space) under the constraint that no two points are closer than $2r$ where $r$ is the ``hard disk'' radius. Instead of $r$, we use the normalized hard disk packing fraction $\rho\in[0,1]$, which is maximal for a hexagonal lattice.  At $\rho\sim 0.72$, the hard disk process undergoes a phase transition that destroys the lattice order. The limit $\rho=0$ corresponds to a Poisson point process.
To sample from a hard disk point process, we use the event-chain Monte Carlo algorithm of Ref.~\cite{Bernard.etal2009}, with code based on that provided with Ref.~\cite{Li.etal2022}. Note that with other algorithms such as sequential deposition or standard Monte Carlo it is either impossible or prohibitively slow to sample at arbitrary values of $\rho$.

To create a tension triangulation, we then compute the Delaunay triangulation of this point set (we further refine the initial condition by removing the edge of this tension triangulation, where the Delaunay algorithm can create very long tension edges). This triangulation is stretched uniformly to set up the initial anisotropy of tension. Cell numbers and initial aspect ratios were also chosen identically in the simulations for the phase diagram Fig.~\ref{fig:tissue-model-disorder}D.

\paragraph*{Boundary conditions, tissue patterning, and isogonal stretching.}

For the first half of the work, Figs.~\ref{fig:tissue-model-disorder}--\ref{fig:tension-config-data}, we use a tissue made of identical cells with free boundary conditions. We now specify the boundary conditions, and the division of the tissue into active and passive patches of the simulations for Figs.~\ref{fig:tissue-model-WT}--\ref{fig:tissue-model-mutants}.

We use appropriate boundary conditions to mimic the geometry of the \textit{Drosophila} lateral ectoderm. We consider an initially rectangular tissue patch with height $>$ width. The top/bottom edges correspond to the dorsal pole, the middle to the ventral region, and the left/right sides to the anterior/posterior edges of the lateral ectoderm. To simulate the cylindrical geometry of the \textit{Drosophila} trunk, the top and bottom boundaries are given slip-wall boundary conditions, i.e.\ the cells on the boundary are pinned along the $y$-axis and free to move along the $x$-axis.
The left/right boundaries are free. Boundary conditions are numerically implemented by a confining potential which penalizes deviation of the $x$ or $y$ coordinate of the cell centroids designated to be boundary cells from the location of the slip wall. This allows for flexible implementation of various boundaries and geometries.
We can also choose different boundary conditions. We can also add slip walls on the left/right boundaries to model the experiments carried out in Ref.~\cite{Collinet.etal2015}, where GBE was blocked by cauterization of the ventral tissue close to the posterior pole.

Next, we designate active and passive regions in the tissue. In the active region, tensions evolve in time according to excitable tension feedback \eqref{eq_triangulation_euler}. In the passive regions, tensions are governed by homeostasis (see below), and cells are softer.
To simulate a wild-type embryo, we chose the size of the passive patch to be approximately 40\% of the embryo circumference (the exact value does not have an important influence). To mimic an embryo without modulation of activity along the DV axis (e.g. a \emph{TollRM9} mutant), all cells along the embryo were taken to be active. The cells at the very tissue boundary are always passive for reasons of numerical stability. Otherwise, edges on the boundary could undergo a T1, which leads to an invalid state.

We model a situation where the cells are initially stretched isogonally. We implement this scenario by making the reference cell shape $S_0$ anisotropic, as discussed above. For the simulations shown in Figs.~\ref{fig:tissue-model-WT} and~\ref{fig:tissue-model-mutants}, we set the initial stretching to zero.

\paragraph*{Dynamics of tensions and cell shapes in passive cells.}

To model the cells in passive tissue regions in our simulations, we must prescribe their tension dynamics and their cell elasticity. Regarding the tension dynamics, we assume that it is governed by tension homeostasis. This means that the intrinsic tensions $\tilde{T}_{ij}$ in the passive region behave as:
\begin{align}
    \frac{d\tilde{T}_{ij}}{dt} = -\tau_{h}^{-1} (\tilde{T}_{ij} - \tilde{T}_{ij, 0})
\end{align}
where $\tau_{h}$ is the homeostasis timescale, and $\tilde{T}_{ij, 0}$ is the tension on the edge at the beginning of the simulation. For newly generated edges after a passive T1, we generate a random reference value $\tilde{T}_{ij, 0} \approx 1$. One can also set $\tilde{T}_{ij, 0}\equiv 1$ for all passive cells, in which case the passive tissue relaxes to a hexagonal lattice.
The speed of homeostasis $\tau_{h}$ has no strong influence on the simulations, but must be large enough so that tensions relax after a passive T1 process before a second passive T1 process takes place. In our simulations, we always set $\tau_{h} = \tau_p$, the speed of passive tension relaxation in the active tissue.

Regarding cell shape dynamics, we model the passive tissue as very soft. In particular, we assume that their shear modulus is very low:  $\mu_p = 0.2\mu_a, \; \lambda_p = 0.2\lambda_a$ where the subscripts
$_a$ and $_b$ refer to active and passive cells, respectively. Further, we use a lower value of the angle-constraint penalty $1/\varepsilon$ than in the active region: $1/\varepsilon_{p} = 1/(200\varepsilon_{a})=5$.
This accounts for the fact that in the passive region, the elastic moduli are much lower (so $1/\varepsilon_{p}$ must be decreased to keep the ratio constant), as well as for the fact that in the passive region, we expect weaker tension. Tension is still the dominant force and the cell array angles match the angles prescribed by the triangulation to within $\sim 2^\circ$. However, we find that small deviations from the triangulation are important to facilitate passive T1s.

\subsection*{Simulation statistics and computing resources}

For each quantification shown in the figures of this paper, by default, we pooled or averaged 3 simulation runs with identical parameters while using different random seeds for the creation of the initial tension triangulation. The phase diagram Fig.~\ref{fig:tissue-model-disorder}D shows the average of 6 simulation runs each. The time traces in Fig.~\ref{SI-fig:cauterized}B as well as the interface length and orientation statistics in Fig.~\ref{SI-fig:cauterized}C were computed from a single simulation. 

Unless specified otherwise, we simulated cell arrays with $N_\mathrm{cells}\approx10^3$ cells (the exact number could vary slightly due to randomness in the hard disk point process). Simulations of saturating feedback in Fig.~\ref{fig:saturating-feedback} were performed with $N_\mathrm{cells}\approx350$ cells, and simulations in Figs.~\ref{fig:tissue-model-WT}--\ref{fig:tissue-model-mutants} had $N_\mathrm{cells}\approx750$ cells.
Simulation edges were always excluded from the analysis. For the single-triangle simulation in Fig.~\ref{fig:tension-config-data}, we simulated $4\times10^3$ individual triangles from the initial condition of the full numerical simulations.
When showing time traces, we smoothed the T1-rate and aspect ratio time series with a Gaussian kernel of width $10dt$ where $dt=\tau_T/200$ is the simulation time step. Numerical parameters were adjusted so that elastic energy minimization and triangulation flattening converged for all simulations and all time points.

Simulations were carried out on a Supermicro workstation with 64 Intel Xeon Gold 6326 cores and 256GB of RAM. However, individual simulations (rather than parameter scans and phase diagrams) can be run without problem on a consumer-grade laptop.

\subsection*{Model parameters and their effects}

As a summary, Table \ref{table:simulation_defaults} gives a list of all the parameters of the tissue-scale model displayed in this work, as well as their default values.
Fig.~\ref{SI-fig:parameter_scan} shows analyses of simulations where several of the parameters are varied, confirming that the qualitative behavior of the model is largely independent of the exact parameter values.

\section{Movie Captions}

\textbf{Tissue shape change by dynamics tension constraints.} \textnormal{Implementation of our model for a symmetric, regular cell array, characterized by one angle $\phi$, determined by the tensions $T_0, T_1$, and two lengths, $\ell_0$ and $\ell_1$, parametrizing the soft isogonal modes. The contour plot shows the cell shape energy $E_\mathcal{C}$ in the incompressible limit where $\ell_1$ is determined uniquely by $\ell_0$ and $\phi$. 
Relaxation of the sub-dominant cell shape energy $E_\mathcal{C}$ is constrained to the isogonal subspace (white line) determined by tension force balance. For the critical tension ratio $T_0/T_1 = \sqrt{2}$, constraining the angle to $\phi_c = \pi/2$ (black dashed line), the interface length minimizing $E_\mathcal{C}$ vanishes (red half-disk).}

\textbf{Simulations comparing ordered and disordered initial cell packings.}
\textnormal{The total extent of tissue shape change depends on the initial degree of order in the cell packing (cf.\ Fig.~\ref{fig:tissue-model-disorder}). Starting from a hexagonal packing, cell rearrangements are initially well-coordinated leading to coherent tissue flow. As order is lost, tissue flow stalls.
The number of cells and the initial tension anisotropy, as well as all other parameters, are the same for both simulations. The simulations were carried out with free boundary conditions.
The top row shows the cell arrays, colored by cell coordination number (hexagons=white), bottom row the underlying tension triangulation.}

\textbf{Simulations comparing saturating and winner-takes-all tension feedback.}
\textnormal{Comparison of saturating (left) and winner-takes-all (right) tension feedback (cf.\ Fig.~\ref{fig:saturating-feedback}). 
Both simulations were initialized with identical initial conditions (an ordered cell array with 20\% tension anisotropy). All other parameters are the same for both simulations.
The top row shows the cell arrays, colored by cell coordination number (hexagons=white), bottom row the underlying tension triangulation.
Saturating feedback leads to little large-scale shape change and creates tension cables.}

\textbf{Simulation combining active and passive tissue regions.}
\textnormal{The embryo is modeled by a rectangular tissue patch with slip walls on the top and bottom edge, mimicking the embryo's cylindrical geometry (cf.\ Fig.~\ref{fig:tissue-model-WT}). The active and passive tissue regions, indicated by color and shading of the edges (color/solid=active), correspond to the germ band and the amnioserosa, respectively. Left: Cell tessellation with black dots indicating the cells along the slip wall (restricted to move horizontally along the ``AP axis''). Right: underlying tension triangulation. The tension triangulation is initialized with the experimentally observed initial tension anisotropy which breaks rotational symmetry and aligns the tissue extension along the ``AP axis''. Colored dots follow a selected subset of cells to highlight rearrangements.}

\textbf{Movie 5: Simulation of convergent extension without cell rigidity and without DV modulation of activity.}
\textnormal{Simulation of tissue scale model of convergent extension in an embryo-like geometry for a control and two ``mutant'' configurations, corresponding 
Fig.~\ref{fig:tissue-model-mutants}.
Left: ``Control'', as in Movie 3. Middle: Active tissue with vanishing shear modulus $\mu_a=0$. Right: No DV modulation, i.e.\ no soft, passive region.
All other parameters are identical. The top shows the cell array, bottom underlying tension triangulation. Active and passive tissue regions are indicated by color and shading of the edges (color/solid=active). 
In all three conditions, active T1s drive elongation of the tension triangulation, while the convergent extension of the cell array only occurs for the ``control'' condition.}

\begin{figure*}[tp]
    \centering
    \includegraphics{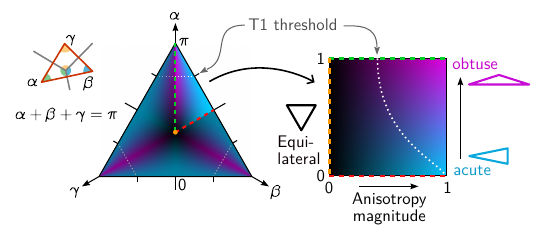}
    \caption{
    Illustration of LTC parameter space in terms of internal triangle angles $(\alpha,\beta,\gamma)$. These angles lie in a simplex defined by $\alpha,\beta,\gamma\geq 0$ and $\alpha+\beta+\gamma=\pi$. 
    The simplex has a 6-fold symmetry due to permutations of the angles; its fundamental domain is mapped into the LTC space spanned by $\tilde{s},\tilde{\psi}$ defined in \eqref{eq:SVD}. The Delaunay threshold $\max \{\alpha,\beta,\gamma\} \leq \pi$ is indicated as a white dashed line.
    }
    \label{SI-fig:triangle_shape_space}
\end{figure*}

\begin{figure}[tp]
    \centering
    \includegraphics{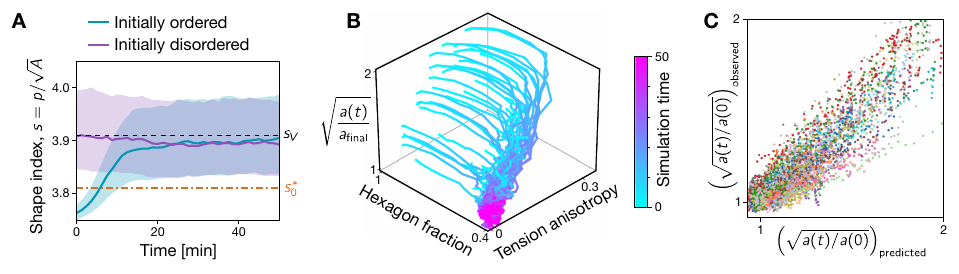}
    \caption{Additional quantifications of cell-packing order and tissue deformation.
    \textbf{A}~As cell-packing order is lost due to rearrangements, the cell-shape index $s$ converges to that of a random Voronoi tessellation $s_\mathrm{V}$ (dashed black line). Note that the critical shape index for the rigidity transition in the ``classical'' vertex model \cite{Bi.etal2015}, $s_0^* \approx 3.81$ (dot-dashed orange line), does not feature in our results. Lines show the median, shaded bands indicate quartiles.
    \textbf{B}~Hexagon fraction, $p_6$, and tension anisotropy $||\langle \sqrt{2} \tilde{Q} \rangle||$ decrease as cells rearrange and eventually converge to a fixed point where anisotropy vanishes and the hexagon fraction matches a random Voronoi tessellation ($\sim 0.4$). At this point, convergent-extension flow, as measured by the remaining change in tissue aspect ratio $a_\mathrm{final}/a(t)$, stalls (cf.\ Figs.~\ref{fig:tissue-model-disorder}C and~\ref{fig:tissue-model-disorder}E). The trajectories approximately collapse onto a plane, suggesting that a linear relationship can predict the convergent extension in terms of $p_6$ and $||\langle\sqrt{2} \tilde{Q} \rangle||$; see \eqref{eq:linear_fit_extension}.
    \textbf{C}~Correlation between predicted and observed change in aspect ratio. Data points are from simulations with initial conditions, spanning the entire parameter diagram in (A), showing that the prediction works across many different initial conditions.
    Points with the same color indicate different time points from the same simulation.
    }
    \label{SI-fig:disorder}
\end{figure}

\begin{figure*}[tp]
    \centering
    \includegraphics[width=\textwidth]{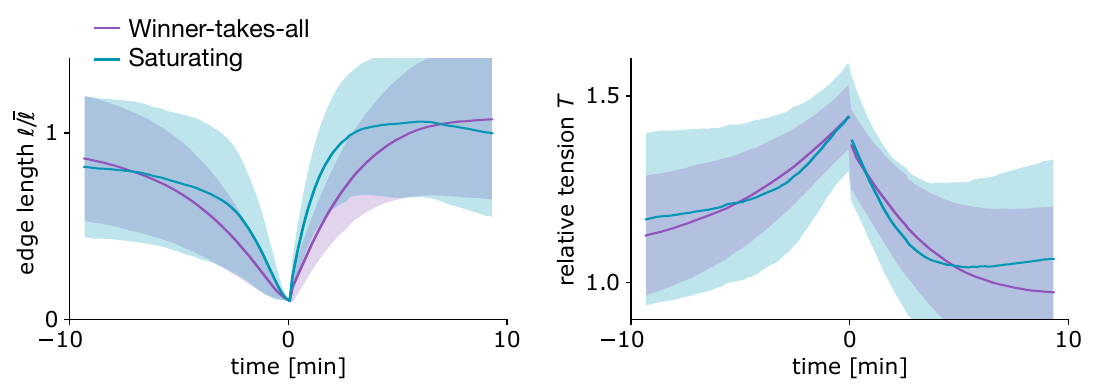}
    \caption{
    Tension and length dynamics of intercalating interface for T1 events in winner-takes-all and saturating feedback simulations. The shaded area indicates standard deviations. Time $t=0\mathrm{min}$ indicates the moment of the T1 transition. Winner-takes-all feedback matches the corresponding experimental data well (see the companion paper \cite{Brauns.etal2024elife}), while saturating feedback does not, featuring a rapid instead of gradual collapse and reemergence of interfaces.
    Interface length is measured relative to the average interface length $\overline{\ell}$.
    T1s occur at a non-zero length due to a numerical cutoff.
    }
    \label{SI-fig:T1-dynamics}
\end{figure*}

\begin{figure*}[tp]
    \centering
    \includegraphics[width=\textwidth]{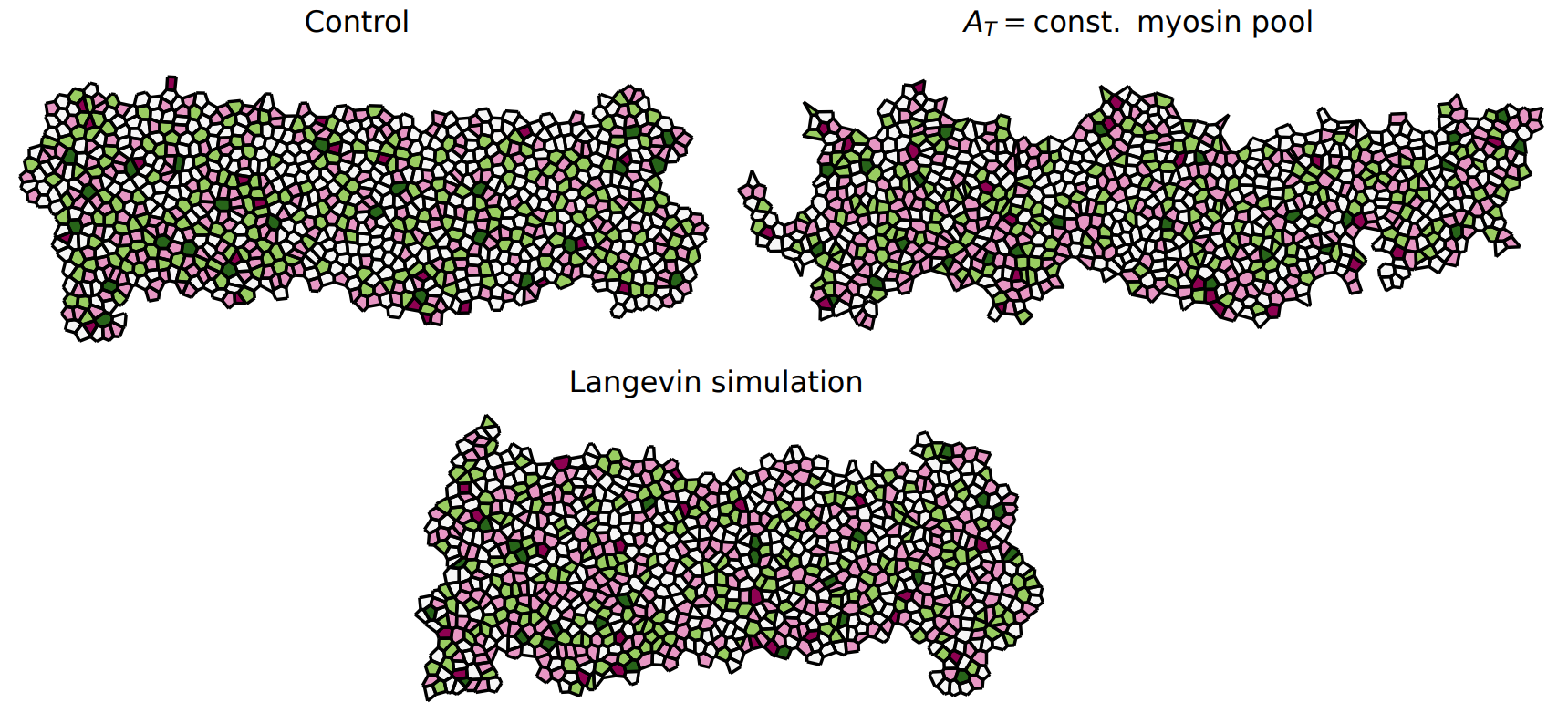}
    \caption{Final tissue shapes (40~min simulation time) for the alternative tissue dynamics models discussed in Sec.~\ref{SI:additional-simulations}D. Simulations were initialized with an ``ordered'' tension triangulation and initial anisotropy $s=0.2$. Cells are colored by coordination number (same color code as in Fig.~\ref{fig:tissue-model-disorder}A).
    }
    \label{SI-fig:alternative_models_meshes}
\end{figure*}

\begin{figure*}[tp]
    \centering
    \includegraphics[width=0.8\textwidth]{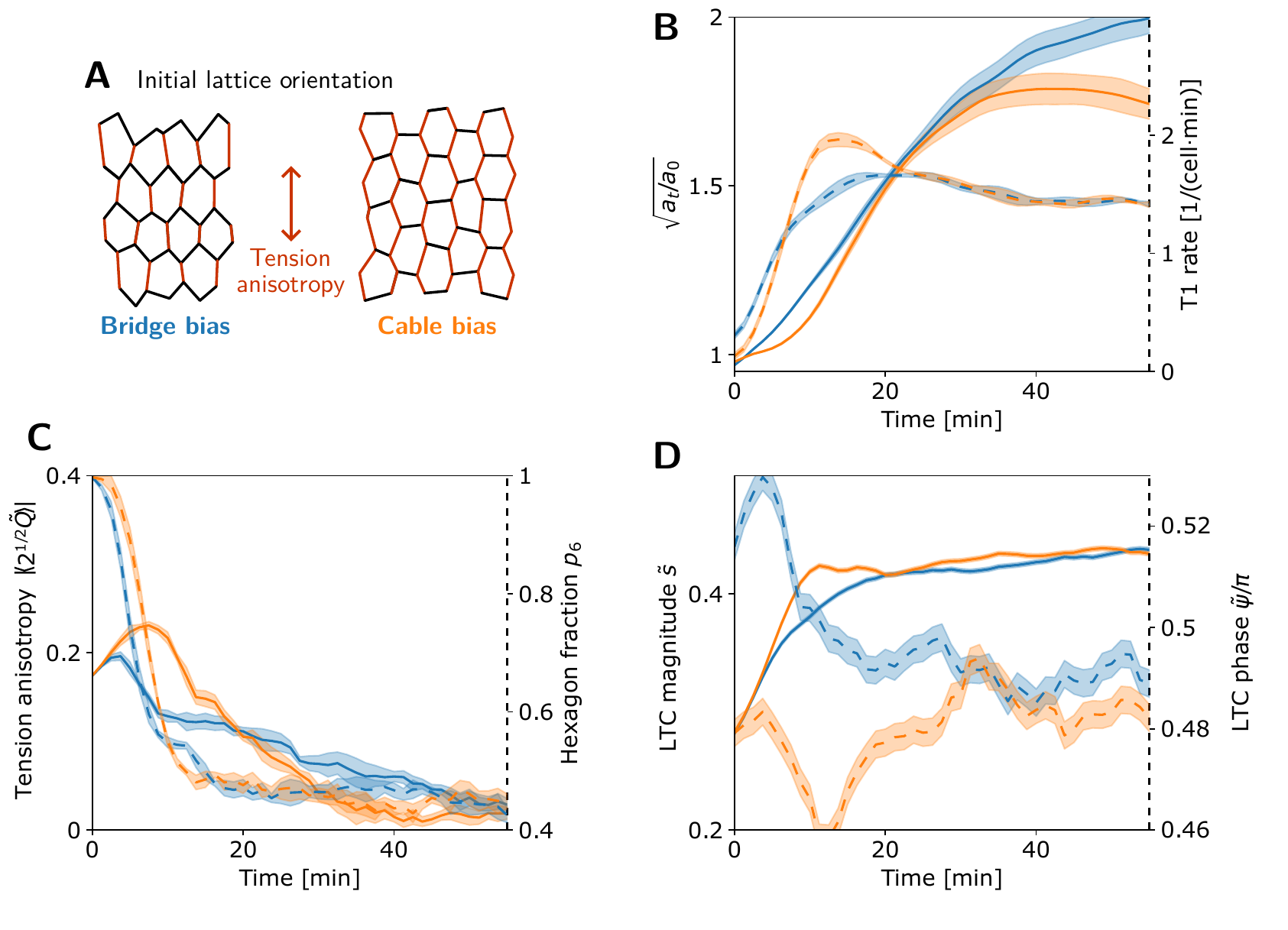}
    \caption{
    Dependence of tissue dynamics on initial tension bridge vs tension cable bias.
    \textbf{A}~The orientation of a hexagonal array of cells relative to the axis of tension anisotropy determines whether the anisotropic tensions manifest primarily as tension bridges (left) or tension cables (right). 
    \textbf{B--D}~Quantification of simulations initialized with hexagonal cell arrays with bridge bias (blue curves) and cable bias (orange curves).
    \textbf{B}~Change of tissue aspect ratio (solid lines, left axis) and T1 rate (dashed lines, right axis) show that convergent extension is faster and more efficient (fewer T1s are required for the same shape change) for an initial bridge bias. Shaded bands indicate standard deviation.
    \textbf{C}~Magnitude of mean tension anisotropy (solid lines, left axis) and hexagon fraction (dashed lines, right axis). Shaded bands indicate standard deviation.
    \textbf{D}~Mean local tension anisotropy magnitude (solid lines, left axis) and median LTC phase (dashed lines, right axis) showing the initial bridge/cable bias and eventual convergence of the LTC phase. Shaded bands indicate standard error. Initial tension anisotropy magnitude $s=0.2$.
    }
    \label{SI-fig:cable_vs_bridge_initial}
\end{figure*}

\begin{figure*}[p]
    \centering
    \includegraphics[width=\textwidth]{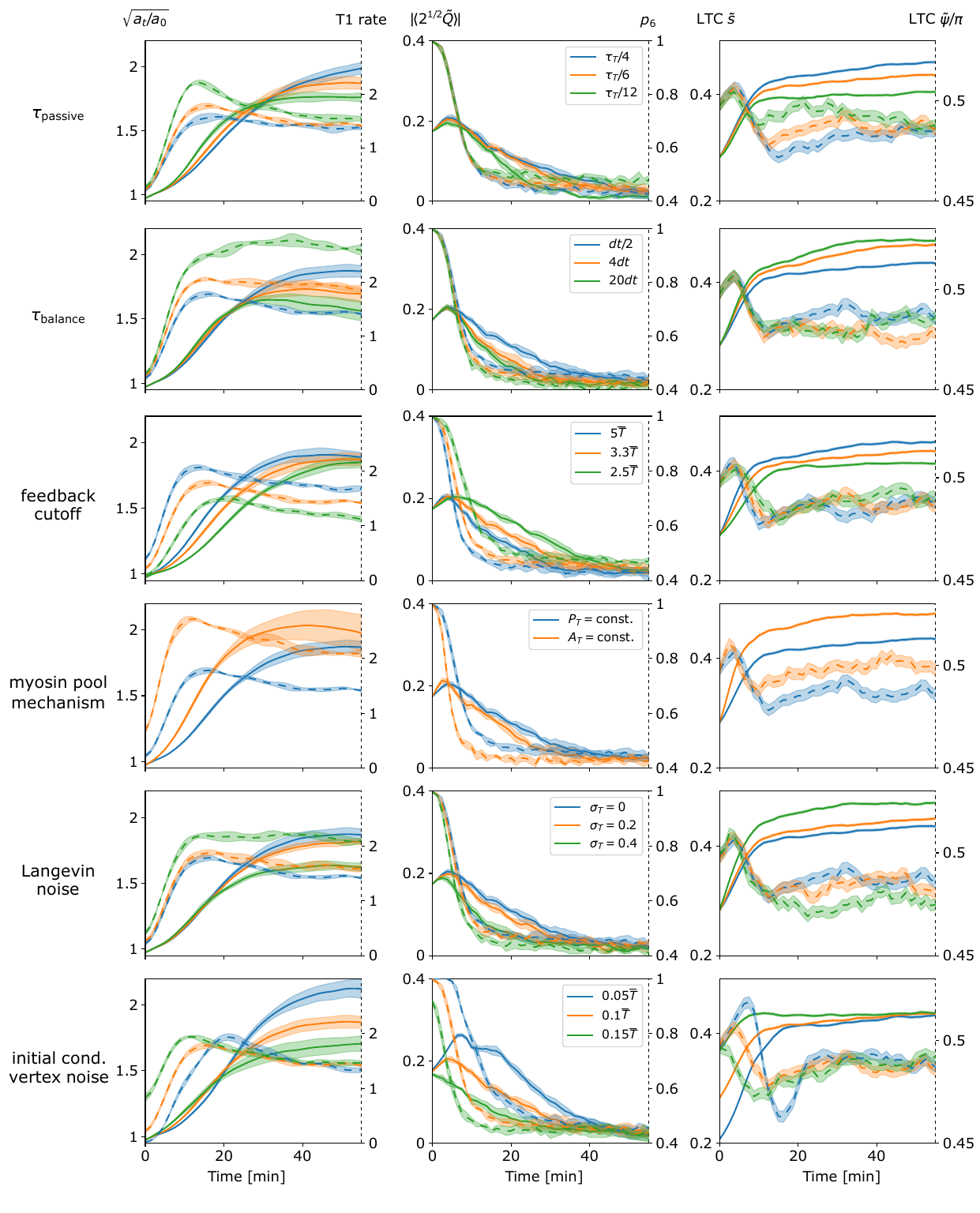}
    \caption{
    Dependence of model behavior on parameters. 
    Each row of plots shows the dependence on one parameter or modeling choice. The three columns follow the same convention as Fig.~\ref{SI-fig:cable_vs_bridge_initial}B--D.
    The passive relaxation time $\tau_p$, the balancing time $\tau_\mathrm{balance}$, the feedback cutoff, and the initial condition vertex noise are explained in more detail in SI Sec.~\ref{SI:simulation_methods}.
    }
    \label{SI-fig:parameter_scan}
\end{figure*}

\begin{figure*}[tp]
    \centering
    \includegraphics[width=0.8\textwidth]{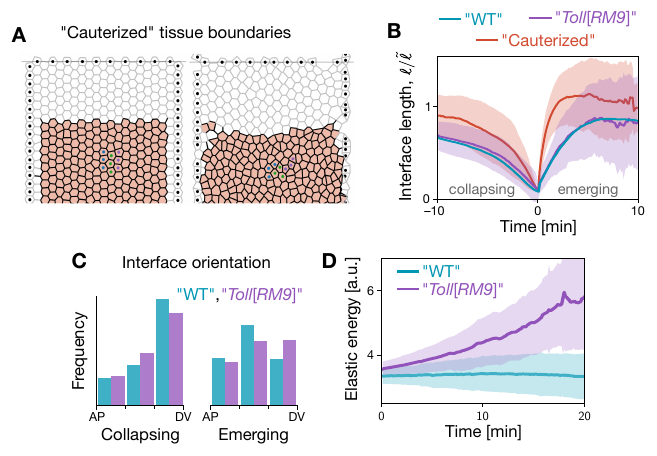}
    \caption{
    Simulations with blocked tissue extension.
    \textbf{A}~Snapshots from a simulation where tissue extension is blocked by slip walls along the AP boundaries of the tissue patch. This mimics experiments where the germ band is blocked from extending via cauterization of the ventral tissue close to the posterior pole, or mutations in the posterior midgut \cite{Collinet.etal2015}.
    \textbf{B}~In simulations where net tissue deformation is blocked (``Toll[RM9]'', ``Cauterized''), the interface length dynamics during active T1s is qualitatively similar to simulations of a freely deforming tissue (``WT'').
    \textbf{C}~Orientation of collapsing and emerging interfaces showing that only the orientation of the latter is qualitatively affected by tissue-scale modulation of activity. The orientation of emerging interfaces is slightly biased along the DV direction in the absence of a passive tissue (``Toll[RM9]'', see Fig.~\ref{fig:tissue-model-mutants}B for snapshots), while it is biased in AP direction in the ``WT'' case (see Fig.~\ref{fig:tissue-model-WT}C for snapshots).
    \textbf{D}~When tissue extension is blocked (``Toll[RM9]'') cell rearrangements are compensated by cell elongation leading to a build-up of elastic energy.
    }
    \label{SI-fig:cauterized}
\end{figure*}

\begin{figure*}[tp]
    \centering
    \includegraphics[width=0.9\textwidth]{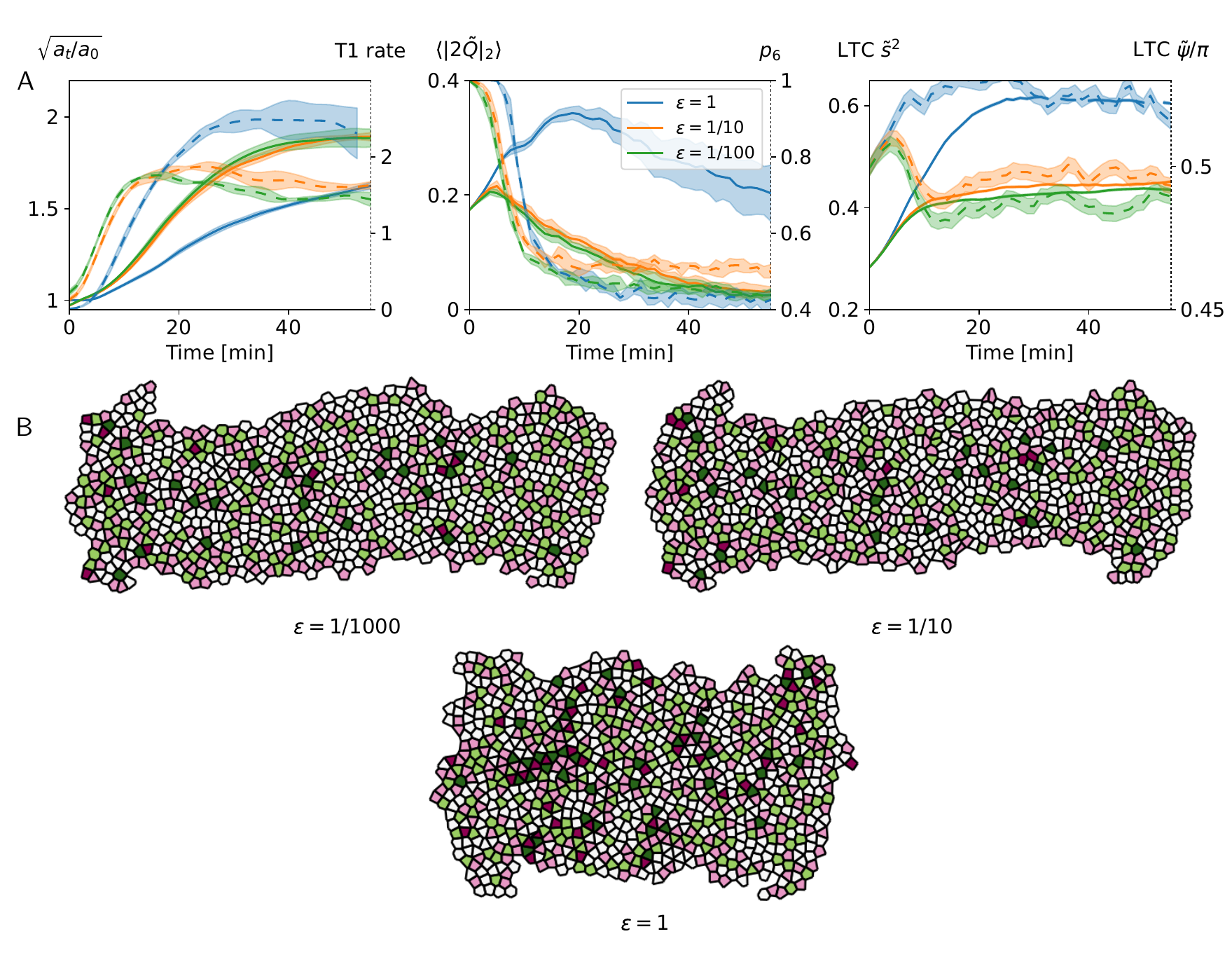}
    \caption{
    Simulations for different values of the tension-isogonal scale-separation parameter $\varepsilon$. 
    \textbf{A} Quantification of tissue deformation and order parameters. Note that the change of aspect ratio is nearly identical for $\varepsilon=0.001, 0.1$ and $1$. 
    \textbf{B} Snapshots at time $t=30\mathrm{min}$ for $\varepsilon=0.001$, $\varepsilon=0.1$, and $\varepsilon=1$.}
    \label{SI-fig:epsilon_dependence}
\end{figure*}

\begin{figure*}[tp]
    \centering
    \includegraphics[width=0.4\textwidth]{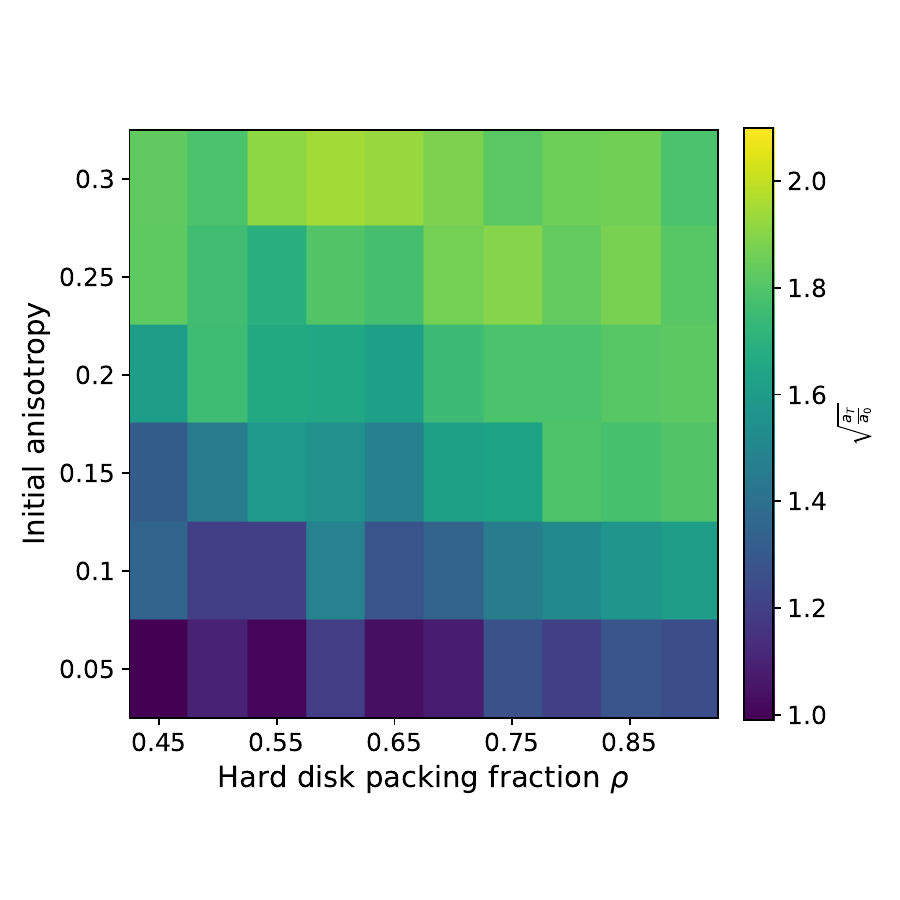}
    \caption{
    Dependence of convergent extension on initial order and anisotropy for vanishing shear rigidity. Heatmap of tissue elongation (square root of aspect ratio change) for vanishing cell shear rigidity, $\mu=0$.
    (Compare to Fig.~\ref{fig:tissue-model-disorder}D where $\mu=1$ was used.)
    For lower shear rigidity, less convergent extension takes place, as shown in Fig.~\ref{fig:tissue-model-mutants}F. The overall structure of the phase diagram remains similar, with weaker dependence on initial order. 
    Note that the simulation was performed with free boundary conditions, i.e.\ there is no external resistance to tissue deformation. An adjacent passive tissue with non-zero shear modulus would suppress convergent extension further (cf.\ Fig.~\ref{fig:tissue-model-mutants}E).
    }
    \label{SI-fig:phase_diagram_shear_modulus}
\end{figure*}

\begin{figure*}[tp]
    \centering
    \includegraphics{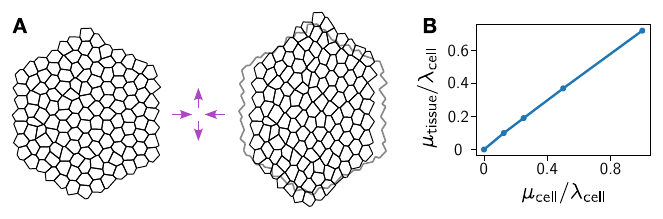}
    \caption{
    \textbf{Cell-level shear modulus determines tissue-level shear modulus.}
    \textbf{A} Large-scale tissue shears are the lowest energy isogonal modes. Reference configuration (minimum energy for a given tension triangulation, left) and deformed configuration (right, outline shows reference configuration), deformed according to the isogonal Hessian eigenvector of one of the two lowest eigenvalues in an example tissue patch.
    \textbf{B} Tissue level ``isogonal shear modulus'' (see \eqref{eq:isogonal_shear_modulus} for a definition) is proportional to cell level shear modulus. Tissue level isogonal shear modulus vs cell-level shear modulus for the tissue patch shown in A.
    }
    \label{SI-fig:isogonal-shear}
\end{figure*}

\begin{figure*}[tp]
    \centering
    \includegraphics[width=0.8\textwidth]{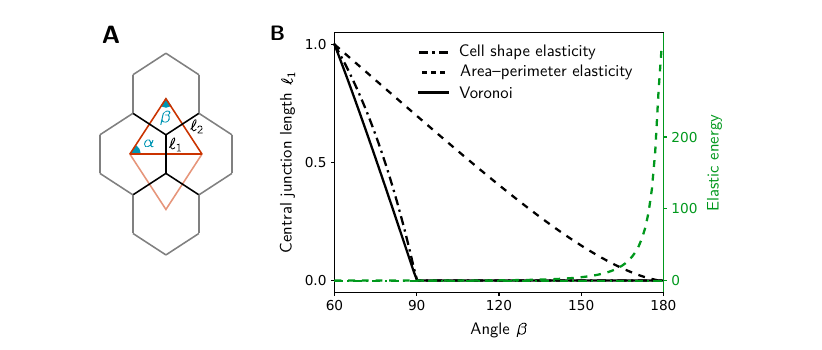}
    \caption{
    Absence of energy barrier to intercalations
    \textbf{A}~Length of the central interface in a quartet for a prescribed isosceles tension triangle parametrized by the angle $\beta$ (the isosceles property implies $\alpha = (\pi - \beta)/2$). 
    \textbf{B}~The solid line shows the interface length in the Voronoi tessellation $\ell_\mathrm{ref}$ (cf.\ \eqref{eq:Voronoi-length}).
    The interface length obtained by minimizing elastic energy \eqref{eq:cell_shape_energy} with isotropic target shape (dot-dashed line) closely follows the Voronoi length and vanishes at the same critical angle.
    By contrast, minimizing the ``area--perimeter'' energy of the vertex-model $E_\mathrm{cell} \sim (A-A_0)^2 + (P-P_0)^2$ in the solid phase, $P_0/\sqrt{A_0} < 3.81$, gives an interface length (dashed black line) that vanishes only for $\beta \to \pi$ and the energy diverges in this limit (dashed green line).
    }
    \label{SI-fig:quartet-model-energy-barrier}
\end{figure*}

\begin{figure*}[p]
    \begin{tcolorbox}{\textbf{Initial condition creation}}
    \begin{enumerate}
        \item Place tension vertices $\mathbf{t}_i$ (e.g. by sampling from a hard-disk process).
        \item Compute Delaunay triangulation of tension vertices.
        \item Apply shear transformation to determine initial anisotropy:
        
         $\mathbf{t}_i \mapsto \mathrm{diag}(\sqrt{1 + s}, 1/\sqrt{1 + s}) \, \mathbf{t}_i$.
         \item Initialize intrinsic tensions $\tilde{T}_{ij}=|\mathbf{t}_i-\mathbf{t}_j|$ and $T_{p, ij}=0$.
         \item Initialize cell vertices $\mathrm{r}_{ijk}$ as Voronoi dual of the Delaunay triangulation.
    \end{enumerate}
    \end{tcolorbox}

    \begin{tcolorbox}{\textbf{Time stepping}}

    \vspace{.125cm}
    Choose time-step size $dt$. For $i=1,...,N_\mathrm{steps}$ do:
    \begin{enumerate}
        \item Update intrinsic tensions (positive feedback Euler step):
        
        $\tilde{T}_{ij} \mapsto \tilde{T}_{ij} + f(\tilde{T}_{ij},...)dt$. $f$ is the feedback RHS.
        \item Update tension triangulation vertices (least-squares fit to intrinsic tensions):
        
        $\mathbf{t}_i \mapsto \mathrm{argmin} \frac{1}{2N_\mathrm{edges}} \sum_{ij} (|\mathbf{t}_i-\mathbf{t}_j| - \tilde{T}_{ij})^2 + \frac{1}{2N_\mathrm{tri}} \ \sum_{ijk} \mathrm{Pen}(A_{ijk}, A_0)$
        \item Update intrinsic tensions (flatness feedback Euler step): 
        
        $\tilde{T}_{ij} \mapsto \tilde{T}_{ij} - \tau_\mathrm{balance}^{-1}  (|\mathbf{t}_i-\mathbf{t}_j| - \tilde{T}_{ij}) dt$ 
        \item Update cell vertices by angle-constrained energy minimization: 
        
        $\mathbf{r}_{ijk} \mapsto \mathrm{argmin} \frac{1}{2N_\mathrm{tri}} \sum_{ij} T_{ij} \overline{\ell} (1 - \hat{n}_{ij} \cdot \mathbf{r}_{ij} )  + \frac{\varepsilon}{2N_\mathrm{cells}} \sum_i E_{\mathcal{C}, i}(S_i)$
        \item Update topology: flip all edges of length $\ell_{ij} < \ell_\mathrm{min}$, update passive tensions $T_{p,ij}$ and intrinsic tensions $\tilde{T}_{ij}$ in flipped edges.
        \item Log: save topology, tension vertices, cell vertices, and T1-transitions.
    \end{enumerate}
    \end{tcolorbox}
    \caption{Simulation workflow showing a summary of the simulation code.}
    \label{SI-fig:simulation_flow_chart}
\end{figure*}

\begin{table}[p]
\centering
\renewcommand{\arraystretch}{1.5}
\begin{tabular}{L{5cm}ccL{8cm}} 
    \toprule
    \textbf{Model parameter} & Symbol & Default value &  Parameter effect, comments \\
    \midrule
    Tension time scale & $\tau_T$ & $25\;\mathrm{min}$ & Fitted to experimental T1 tension dynamics.  \\
    Feedback exponent & $n$ & $4$ & Must be $>0$ to produce flow. Weak effect on net CE.  \\
    Tension balancing & $\tau_\mathrm{balance}$  & $\tau_T /200$ &  Must be $\ll \tau_T $. Higher values lead to a higher rate of disoriented T1s.
    \\
    Passive tension relaxation & $\tau_p$  & $\tau_T/6$ & No strong effect. Must be sufficiently small $< \tau_T/2$ to create irreversible T1s.
    \\
    Elasticity coefficients & $\lambda, \mu, \lambda_a, \mu_a$  & $1$  & No strong effect. Must be $>0$\\
    Passive region shear moduli & $\mu_p, \: \lambda_p$ &  $0.2$ 
    & Higher shear resistance of passive tissue lowers the net amount of CE. \\
    \toprule
    \textbf{Initial condition parameters} & \\
    \midrule
    Average tension & $\overline{T}$ & $1$ &  \\
    Average edge length & $\ell_0$ & $1/\sqrt{3}$ & Set to match the Voronoi dual of the tension triangulation \\
    Initial anisotropy & $||\langle \sqrt{2}\tilde{Q}\rangle||$ & $0.2$ & Higher values increase net CE. \\
    Tension vertex hard disk packing fraction & $\rho$ &  $0.45-0.9$ & Higher values increase net CE. \\
    Gaussian noise in initial tensions (standard deviation) & &  $0.1 \:\overline{T}$ & Weak effect on net CE, as long as noise is sufficiently weak to not affect triangulation topology. \\
    Cell reference shape & $S_0$ & $3\ell_0\: \mathbb{I}$ & Anisotropic $S_0$ increases bridge bias by shifting T1 threshold.  \\
    Number of cells & $N_\mathrm{cells}$ & $\approx 10^3$ & Number can vary due to random sampling. For ``germ band'' simulations we used $N_\mathrm{cells}=750$. \\
    \toprule
    \textbf{Numerical parameters} & \\
    \midrule
    Tension feedback cutoff & $k_\mathrm{cutoff}$ & $0.3\overline{T}$ \\
    Triangulation area penalty & $\gamma_\text{tri}$ & $0.01$ \\
    Angle constraint penalty & $1/\varepsilon$ & $10^3$ & In the passive region, we use $1/\varepsilon=5$ to account for the reduced cortical tensions there. \\
    Boundary condition penalty & & $5\times 10^3$ & {} \\
    Time step & $dt$ & $5\times 10^{-3} \: \tau_T$ \\
    Edge length to trigger collapse & & $0.1 \: \ell_0$ \\
    \bottomrule
\end{tabular}
\caption{\textnormal{Parameters of tissue scale simulation. A value in the second-to-last column indicates a ``default'' value which was used in all simulations except otherwise indicated. $\overline{T}$ is the overall tension scale (average edge tension at the initial condition), and $\ell_0$ is the overall length scale (average edge length at the initial condition).
Net CE denotes the amount of convergent extension as tissue flow saturates.}
}
\label{table:simulation_defaults}
\end{table}

\end{document}